\documentclass[reprint,superscriptaddress,showpacs]{revtex4-1}
\usepackage[pdftex]{graphicx}
\usepackage[title]{appendix}
\usepackage{epstopdf}
\usepackage{caption}
\usepackage{multirow}
\usepackage{varioref}
\usepackage{hyperref}
\usepackage[table]{xcolor}
\definecolor{Gray}{gray}{0.95}

\graphicspath{ {./Figures/} }

\begin{document}

\title[Results and Analysis of LPF in-Flight Charge Control]{Precision Charge Control for Isolated Free-Falling Test Masses: LISA Pathfinder Results}

\def\addressa{European Space Astronomy Centre, European Space Agency, Villanueva de la Ca\~{n}ada, 28692 Madrid, Spain}
\def\addressb{Albert-Einstein-Institut, Max-Planck-Institut f\"ur Gravitationsphysik und Leibniz Universit\"at Hannover, Callinstra{\ss}e 38, 30167 Hannover, Germany}
\def\addressc{APC, Univ Paris Diderot, CNRS/IN2P3, CEA/lrfu, Obs de Paris, Sorbonne Paris Cit\'e, France}
\def\addressd{High Energy Physics Group, Physics Department, Imperial College London, Blackett Laboratory, Prince Consort Road, London, SW7 2BW, UK }
\def\addresse{Dipartimento di Fisica, Universit\`a di Roma ``Tor Vergata'', and INFN, sezione Roma Tor Vergata, I-00133 Roma, Italy}
\def\addressf{Department of Industrial Engineering, University of Trento, via Sommarive 9, 38123 Trento, and Trento Institute for Fundamental Physics and Application / INFN}
\def\addressh{European Space Technology Centre, European Space Agency, Keplerlaan 1, 2200 AG Noordwijk, The Netherlands}
\def\addressi{Dipartimento di Fisica, Universit\`a di Trento and Trento Institute for Fundamental Physics and Application / INFN, 38123 Povo, Trento, Italy}
\def\addressj{The School of Physics and Astronomy, University of Birmingham, Birmingham, UK}
\def\addressl{Institut f\"ur Geophysik, ETH Z\"urich, Sonneggstrasse 5, CH-8092, Z\"urich, Switzerland}
\def\addressm{The UK Astronomy Technology Centre, Royal Observatory, Edinburgh, Blackford Hill, Edinburgh, EH9 3HJ, UK}
\def\addressn{Institut de Ci\`encies de l'Espai (CSIC-IEEC), Campus UAB, Carrer de Can Magrans s/n, 08193 Cerdanyola del Vall\`es, Spain}
\def\addresso{DISPEA, Universit\`a di Urbino ``Carlo Bo'', Via S. Chiara, 27 61029 Urbino/INFN, Italy}
\def\addressp{European Space Operations Centre, European Space Agency, 64293 Darmstadt, Germany }
\def\addressq{Physik Institut, Universit\"at Z\"urich, Winterthurerstrasse 190, CH-8057 Z\"urich, Switzerland}
\def\addressr{SUPA, Institute for Gravitational Research, School of Physics and Astronomy, University of Glasgow, Glasgow, G12 8QQ, UK}
\def\addresss{Department d'Enginyeria Electr\`onica, Universitat Polit\`ecnica de Catalunya, 08034 Barcelona, Spain}
\def\addresst{Institut d'Estudis Espacials de Catalunya (IEEC), C/ Gran Capit\`a 2-4, 08034 Barcelona, Spain}
\def\addressu{Gravitational Astrophysics Lab, NASA Goddard Space Flight Center, 8800 Greenbelt Road, Greenbelt, MD 20771 USA}
\def\addressbb{Department of Mechanical and Aerospace Engineering, MAE-A, P.O. Box 116250, University of Florida, Gainesville, Florida 32611, USA}
\def\addresscc{Istituto di Fotonica e Nanotecnologie, CNR-Fondazione Bruno Kessler, I-38123 Povo, Trento, Italy}
\def\addressdd{OHB Italia SpA, Via Gallarate 150, 20151 Milano, Italy}
\def\addressee{Fakult\"at Elektrotechnik und Informatik, Hochschule Ravensburg-Weingarten, Doggenriedstra{\ss}e, 88250 Weingarten, Germany}
\def\addressff{Airbus Defence and Space, Gunnels Wood Road, Stevenage, Hertfordshire, SG1 2AS, United Kingdom}

\author{M~Armano}\affiliation{\addressh}
\author{H~Audley}\affiliation{\addressb}
\author{J~Baird}\affiliation{\addressd}\affiliation{\addressc}
\author{P~Binetruy}\thanks{Deceased 30 March 2017}\affiliation{\addressc}
\author{M~Born}\affiliation{\addressb}
\author{D~Bortoluzzi}\affiliation{\addressf}
\author{E~Castelli}\affiliation{\addressi}
\author{A~Cavalleri}\affiliation{\addresscc}
\author{A~Cesarini}\affiliation{\addresso}
\author{A\,M~Cruise}\affiliation{\addressj}
\author{K~Danzmann}\affiliation{\addressb}
\author{M~de Deus Silva}\affiliation{\addressa}
\author{I~Diepholz}\affiliation{\addressb}
\author{G~Dixon}\affiliation{\addressj}
\author{R~Dolesi}\affiliation{\addressi}
\author{L~Ferraioli}\affiliation{\addressl}
\author{V~Ferroni}\affiliation{\addressi}
\author{E\,D~Fitzsimons}\affiliation{\addressm}
\author{M~Freschi}\affiliation{\addressa}
\author{L~Gesa}\affiliation{\addressn}
\author{D~Giardini}\affiliation{\addressl}
\author{F~Gibert}\affiliation{\addressi}
\author{R~Giusteri}\affiliation{\addressi}
\author{C~Grimani}\affiliation{\addresso}
\author{J~Grzymisch}\affiliation{\addressh}
\author{I~Harrison}\affiliation{\addressp}
\author{G~Heinzel}\affiliation{\addressb}
\author{M~Hewitson}\affiliation{\addressb}
\author{D~Hollington}\email[Corresponding author: ]{d.hollington07@imperial.ac.uk}\affiliation{\addressd}
\author{D~Hoyland}\affiliation{\addressj}
\author{M~Hueller}\affiliation{\addressi}
\author{H~Inchausp\'e}\affiliation{\addressc}\affiliation{\addressbb}
\author{O~Jennrich}\affiliation{\addressh}
\author{P~Jetzer}\affiliation{\addressq}
\author{N~Karnesis}\affiliation{\addressc}
\author{B~Kaune}\affiliation{\addressb}
\author{N~Korsakova}\affiliation{\addressr}
\author{C\,J~Killow}\affiliation{\addressr}
\author{L~Liu}\affiliation{\addressi}
\author{I~Lloro}\affiliation{\addressn}
\author{J\,A~Lobo}\thanks{Deceased 30 September 2012}\affiliation{\addressn}
\author{J\,P~L\'opez-Zaragoza}\affiliation{\addressn}
\author{R~Maarschalkerweerd}\affiliation{\addressp}
\author{F~Mailland}\affiliation{\addressdd}
\author{D~Mance}\affiliation{\addressl}
\author{V~Mart\'{i}n}\affiliation{\addressn}
\author{L~Martin-Polo}\affiliation{\addressa}
\author{F~Martin-Porqueras}\affiliation{\addressa}
\author{J~Martino}\affiliation{\addressc}
\author{I~Mateos}\affiliation{\addressn}
\author{P\,W~McNamara}\affiliation{\addressh}
\author{J~Mendes}\affiliation{\addressp}
\author{L~Mendes}\affiliation{\addressa}
\author{N~Meshskar}\affiliation{\addressl}
\author{M~Nofrarias}\affiliation{\addressn}
\author{S~Paczkowski}\affiliation{\addressb}
\author{M~Perreur-Lloyd}\affiliation{\addressr}
\author{A~Petiteau}\affiliation{\addressc}
\author{M~Pfeil}\affiliation{\addressd}\affiliation{\addressee}
\author{P~Pivato}\affiliation{\addressi}
\author{E~Plagnol}\affiliation{\addressc}
\author{J~Ramos-Castro}\affiliation{\addresss}
\author{J~Reiche}\affiliation{\addressb}
\author{D\,I~Robertson}\affiliation{\addressr}
\author{F~Rivas}\affiliation{\addressn}
\author{G~Russano}\affiliation{\addressi}
\author{G~Santoruvo}\affiliation{\addressdd}
\author{P~Sarra}\affiliation{\addressdd}
\author{D~Shaul}\affiliation{\addressd}
\author{J~Slutsky}\affiliation{\addressu}
\author{C\,F~Sopuerta}\affiliation{\addressn}
\author{T~Sumner}\affiliation{\addressd}
\author{D~Texier}\affiliation{\addressa}
\author{J\,I~Thorpe}\affiliation{\addressu}
\author{C~Trenkel}\affiliation{\addressd}\affiliation{\addressff}
\author{D~Vetrugno}\affiliation{\addressi}
\author{S~Vitale}\affiliation{\addressi}
\author{G~Wanner}\affiliation{\addressb}
\author{H~Ward}\affiliation{\addressr}
\author{S~Waschke}\affiliation{\addressd}
\author{P\,J~Wass}\affiliation{\addressd}\affiliation{\addressbb}
\author{W\,J~Weber}\affiliation{\addressi}
\author{L~Wissel}\affiliation{\addressb}
\author{A~Wittchen}\affiliation{\addressb}
\author{P~Zweifel}\affiliation{\addressl}

\begin{abstract}

The LISA Pathfinder charge management device was responsible for neutralising the cosmic ray induced electric charge that inevitably accumulated on the free-falling test masses at the heart of the experiment. We present measurements made on ground and in-flight that quantify the performance of this contactless discharge system which was based on photo-emission under UV illumination. In addition, a two-part simulation is described that was developed alongside the hardware. Modelling of the absorbed UV light within the Pathfinder sensor was carried out with the GEANT4 software toolkit and a separate MATLAB charge transfer model calculated the net photocurrent between the test masses and surrounding housing in the presence of AC and DC electric fields. We confront the results of these models with observations and draw conclusions for the design of discharge systems for future experiments like LISA that will also employ free-falling test masses.

\end{abstract}

\maketitle

%%%%%%%%%%%%%%%%%%%%%%%%%%%%%%%%%%%%%%%%%%%%%%%%%%

\section{Introduction}

The space-based gravitational wave observatory, the Laser Interferometer Space Antenna (LISA), will open up a completely new observational window on the universe. By escaping noise sources associated with the terrestrial environment that limit ground based detectors, low-frequency astronomical sources such as merging supermassive black holes, extreme mass-ratio inspirals and thousands of extended compact binary systems will be observed \cite{Amaro2017}. As a technology demonstrator for LISA, LISA Pathfinder was a European Space Agency mission that operated between March 2016 and July 2017. It successfully demonstrated that the residual acceleration between two free-falling test masses could be reduced to fm\,s$^{-2}/\sqrt{\textrm{Hz}}$ levels in a frequency range of 0.02-30\,mHz, essential for a million-km scale gravitational wave observatory \cite{Armano2016, Armano2018a}.

A sub-set of forces contributing to this residual acceleration are related to charge accumulating on the test masses as a consequence of the high-energy charged particles present in the space environment \cite{Shaul2005, Armano2017}. Though the structures surrounding the test masses offer shielding from particles with $<$100\,MeV/nucleon \cite{Araujo2005, Wass2005}, cosmic rays with higher energy inevitably deposit charge. A charged test mass then interacts with the conducting surfaces that surround it via the Coulomb force, or, at a much smaller level, with any residual magnetic fields via the Lorentz force \cite{Jafry1997}. Charging is caused by particle radiation from two main sources: Galactic Cosmic Rays (GCRs) and Solar Energetic Particles (SEPs) \cite{Armano2018b, Armano2018c}. Simulations predicted that the ever present but variable flux of GCRs would charge the test masses positively at a rate of +10 to +100\,e\,s$^{-1}$, while transient SEP events occurring at most a few times per year could enhance the charging rate by several orders of magnitude \cite{Araujo2005}. In addition, a significant net charge could also be transferred to the test masses during their initial release as well as after any subsequent re-grabbing maneuvers. Due to work function differences between surfaces in last direct mechanical contact as well as triboelectric processes, this was predicted to be of order $10^{8}$\,e.

To mitigate the effects of test-mass charging, Pathfinder included a Charge Management System (CMS) as part of the LISA technology package (LTP) instrument. While other sensitive space-borne accelerometers, such as CHAMP \cite{Reigber2002}, GRACE \cite{Tapley2004}, GOCE \cite{Rummel2011} and MICROSCOPE \cite{Touboul2001} have relied on a physical electrical connection to control the charge of the test mass, in order to reach the desired force noise goals for LISA, a contactless method is essential. The Pathfinder CMS exploited the same principles of photo-emission under UV illumination as successfully demonstrated in the Gravity Probe B (GP-B) mission \cite{Buchman1995}. In GP-B, bi-directional charge control was achieved by applying a strong local electric field at the point of illumination. However, in order to allow the possibility for full-precision acceleration measurements to continue during discharging, the Pathfinder system was designed to achieve charge control by differential illumination of the test mass and surrounding housing, avoiding large DC electrostatic fields.

In simple terms, such a system produces positive charging when the test mass is illuminated and photoelectrons are removed, while negative charging occurs when the test mass surroundings are illuminated and photoelectrons migrate to the test mass. However, a real system involves a number of complicating factors. The gold sensor surfaces are highly reflective at the UV wavelengths required for photo-emission and no matter the illumination this results in UV light being distributed on both test mass and housing surfaces. Furthermore, the photoelectron yield can vary significantly due to unavoidable surface contamination during instrument assembly, even between surfaces that were prepared using an identical method. Finally the energy of the liberated electrons are of order 1\,eV so that the Volt-scale AC and DC voltages used for test mass capacitive sensing and actuation can strongly influence the discharging behaviour.

The structure of the paper is as follows. After first describing the Pathfinder CMS hardware and the details of the discharging problem, we present a computer simulation of the system. This model is then used to interpret measurements made on ground and in-flight with a comparison of the underlying surface properties being made. We conclude by discussing the lessons learned that will aid the development of a CMS for LISA.

%%%%%%%%%%%%%%%%%%%%%%%%%%%%%%%%%%%%%%%%%%%%%%%%%%

\section{The LISA Pathfinder Charge Management System}\label{sec:LPF_CMS}

\begin{figure}[]
\centering
\includegraphics[width=0.5\textwidth]{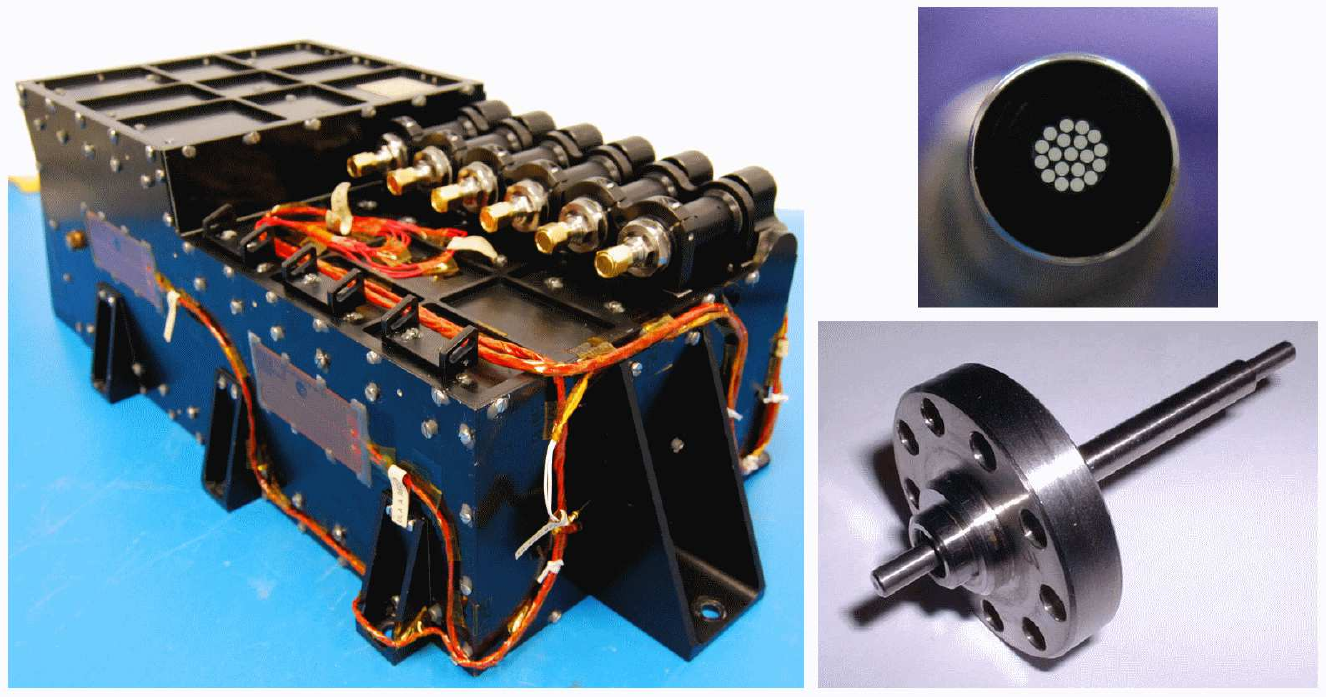}
\caption[CMS Hardware.]{\label{fig:CMS_Hardware} Left: The flight ULU with external heaters attached. The ends of the six optics barrels are seen temporarily sealed with gold dust caps. Top Right: An end-on view of an FOH with the 19 individual cores visible. Bottom Right: A flight ISUK.}
\end{figure}

Delivered by Imperial College London, the Pathfinder CMS consisted of three major hardware components, shown in Fig. \ref{fig:CMS_Hardware}. The UV Lamp Unit (ULU) housed six low-pressure mercury lamps (UVP, model 11SC-2) along with their associated electronics. Such lamps produce a spectrum of emission lines but only the one at 253.7\,nm (4.89\,eV) was used for discharging. A band-pass filter blocked deep UV light to protect optical elements from solarisation while also preventing excess visible photons from entering the sensor. A system of lenses were used to collect and focus the light from an aperture into the Fibre Optic Harness (FOH). The FOH was a series of custom-made, UV-transparent optical fibre assemblies which routed the light from the ULU in the main body of the spacecraft to the two vacuum enclosures containing the gravitational reference sensors (GRS). The routing path for the fibres required two FOHs, joined at the interface of the thermal enclosure of the LTP. At the join, an attenuating spacer was used to adjust the level of light reaching the sensor with the length of each spacer being specific to each lamp. Each FOH assembly consisted of a bundle of 19 individual 200\,$\mu$m-diameter fibres, giving an effective diameter of over 1\,mm but allowing a minimum bend radius of 25\,mm. Finally the Inertial Sensor UV Kit (ISUK) was a custom-made titanium ultra-high vacuum feedthrough with a 1\,mm diameter, 75\,mm long optical fibre that delivered the light into the sensor. Each GRS had three ISUKs that entered at the corners of the lower $z$-face of the sensor housing. Two IUSKs were directed toward the housing and one toward the test mass.

Each lamp housing contained a thermistor, an Ohmic heater and a UV-sensitive silicon carbide photodiode (ifw optronics, model JEC 0.1s) for monitoring the UV power output from the lamp. During on-ground calibration of the instrument, the relationship between each lamp's photodiode reading and the UV power emitted from its ISUK was measured. The photodiode reading was then used in-flight to estimate the photon flux entering the sensor. The UV power entering the sensor was commandable in a range of order 1-100\,nW with 256 output settings.

The Pathfinder test masses were 1.93\,kg, 46\,mm cubes made from 73:27 gold-platinum alloy. Each mass was surrounded by a cuboidal housing made from molybdenum accommodating 12 electrodes for sensing and actuation in six degrees of freedom. The gaps between electrode and test mass ranged from 2.9-4.0\,mm depending on the axis. Test mass actuation was achieved with audio-frequency sinusoidal voltages on the sensing electrodes with maximum amplitudes of up to 12\,V. The $y$ and $z$ housing faces also had an additional six `injection' electrodes, with 4.88\,V applied at 100\,kHz to capacitively bias the test mass at 0.6\,V for position sensing \cite{Weber2003}. The total test mass capacitance was 34.2\,pF and this value will be used throughout the rest of the paper to convert the test mass charge to a test mass potential with respect to the electrically grounded housing.

\begin{figure}[]
\centering
\includegraphics[width=0.5\textwidth]{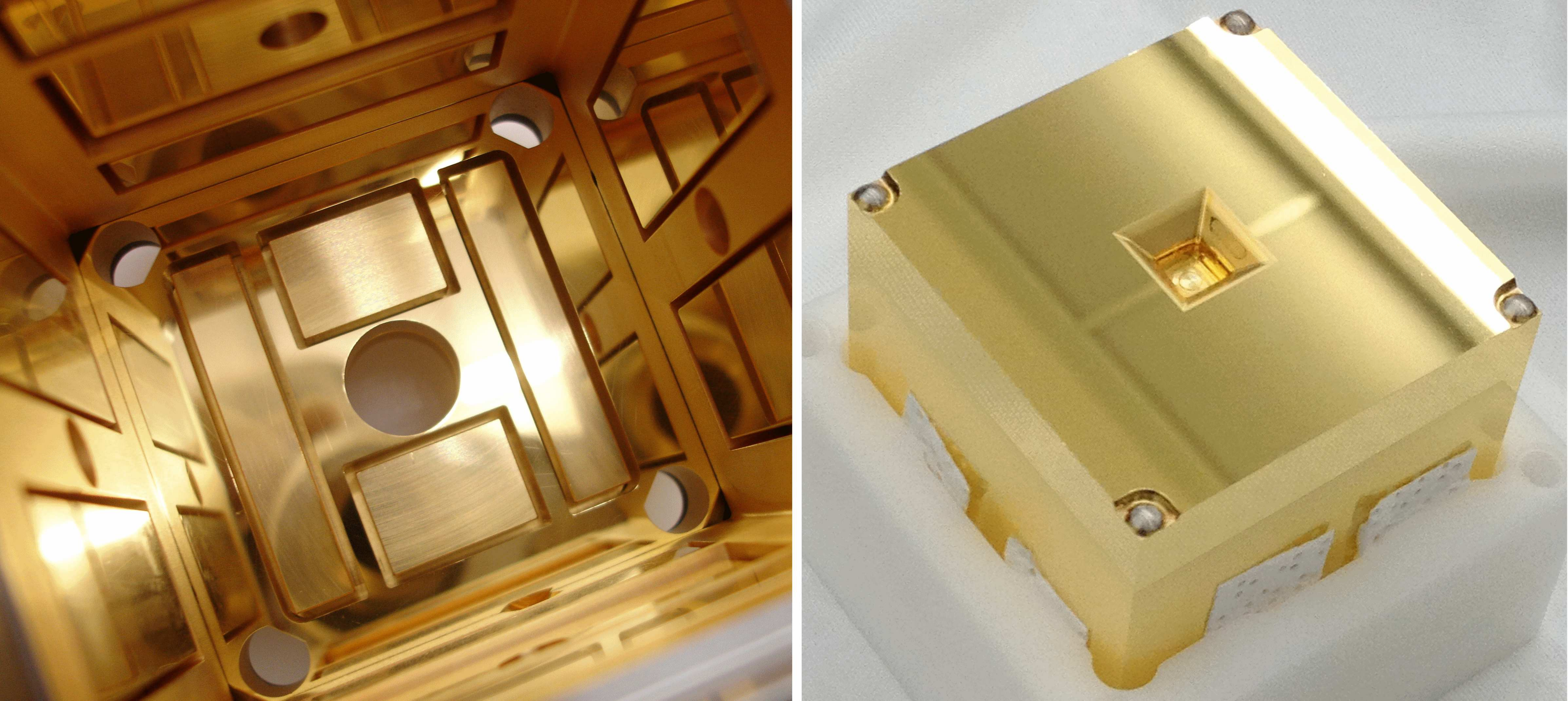}
\caption[GRS Photos.]{\label{fig:GRS_Photos} Left: A view up into the sensor housing, centred on the upper $z$-face. The test mass and lower $z$-face have yet to be integrated and the ISUK entry points are just out of shot. The two $x$-faces are on the left and right of the image while the $y$-faces can be seen at the top and bottom. The empty recesses where four of the caging fingers enter can be seen in the corners and the plunger entry hole can be seen in the centre. The electrodes set within the walls are clearly visible. Right: One of the flight test masses with the lower half temporarily held in a protective case. Note the caging corner domes with the underlying gold-platinum surface exposed. Credit: OHB Italia SpA.}
% Both photos are copyright of OHB Italia SpA but hosted publicly on ESA website. I sent email (cgs@cgspace.it) on 19/04/18 asking for permission to use them. As yet, no reply.
\end{figure}

Almost all of the test mass, housing and electrode surfaces were sputter-coated with gold except for features related to the caging system that held the test masses during launch \cite{Bortoluzzi2009}. Eight iridium caging fingers entered each sensor via holes at the corners of both $z$-faces. A concave hemisphere at the tip of each finger was designed to fit over an adjacent dome on each test mass corner and prevent slippage when the caging system was engaged. Both the fingers and the area where they made mechanical contact with the test mass remained uncoated in order to avoid the risk of the gold pealing during release. Each sensor also had two gold-platinum plungers at the centre of each $z$-face that formed the Grabbing Positioning and Release Mechanism (GPRM) used to release the test masses into free-fall. Once in space the caging fingers were permanently retracted 2.25\,mm below the level of the housing $z$-face with a one-shot mechanism. Upon initial test mass release the GPRM plungers were retracted 6.5\,mm below the housing $z$-face with the plungers returning to the same position following any re-grabbing maneuvers. A flight test mass and electrode housing prior to integration can be seen in Fig. \ref{fig:GRS_Photos}.

To measure the test mass charge, sinusoidally varying voltages at milli-Hertz frequencies were applied to combinations of sensing electrodes. This induced a force or torque on the test mass with an amplitude that was proportional to the test mass charge. Typically this was done in the $x$ or $\phi$ axes using the interferometer to measure the differential test mass motion or rotation with a higher precision than could be achieved with the capacitive sensor \cite{Audley2011}. To limit the coupling to fluctuations in both stray actuation voltages as well as from surface patch effects \cite{Antonucci2012, Armano2017}, the test mass charge needed to be kept below approximately $2\times10^{7}$ elementary charges (equivalent to about $\pm100$\,mV). During the mission this was primarily achieved by employing a so called fast discharge scheme. The test masses were left to charge gradually for pre-determined intervals (typically a couple of weeks) before being discharged relatively quickly (typically tens of minutes). The discharge was performed either by a pre-configured command sequence uploaded from ground or by autonomous onboard software using an iterative sequence of measurement and illumination. An alternative approach, referred to as continuous discharge, was tested on several occasions during the mission. This scheme involved keeping the test mass charge permanently close to zero by continuously illuminating the sensor such that the charging-discharging rates were balanced.

\subsection{Discharge Properties}\label{sec:DischargeProps}

The performance of the Pathfinder CMS depended on the photoelectric properties of the predominantly gold coated GRS surfaces. Deposited in vacuum and measured \textit{in situ}, the work function of pure gold is 5.2\,eV \cite{Huber1966}. However, upon exposure to air, stable adsorbates generally reduce the work function to around 4.2\,eV \cite{Saville1995}. It is therefore only the presence of surface contamination that makes photoemission possible with the 4.89\,eV photons produced by the mercury lamps. The adsorbates are bound to the surface of the gold sufficiently well to persist in high vacuum and with moderate heating \cite{Hechenblaikner2012}. With 4.89\,eV photons, differences in the surface contamination can produce values for the gold quantum yield (the number of photoelectrons emitted per absorbed photon) that vary between 10$^{-6}$ and 10$^{-4}$ \cite{Schulte2009}. Of the remaining exposed surfaces within the sensor, iridium has a work function ranging from 5.4-5.8\,eV \cite{Haynes2010} and should therefore not have contributed to discharging currents in the Pathfinder system. Limited photoelectric measurements of 73:27 gold-platinum suggest properties similar to those of gold \cite{Hollington2011}.

Due to reflections within the sensor, when primary surfaces were illuminated some UV light was inevitably absorbed by opposing surfaces. This resulted in two competing photocurrents, one in the desired direction and one acting against it. For instance, if the test mass was primarily illuminated a photocurrent from the test mass toward the electrode housing was generated but there was also a counter-photocurrent from the housing surfaces toward the test mass. The magnitude of these two photocurrents depended on the amount of light absorbed and the quantum yield of the surfaces. As the photoelectrons had energies of around 1\,eV they were also strongly affected by the Volt-scale AC and DC voltages applied to the electrodes as well as any potential difference between the test mass and electrically grounded housing. 

To assess the discharging behaviour in-flight only the apparent yield could be obtained, which is defined as the net change in test mass charge per photon injected into the sensor. We define a positive apparent yield as that which produced a positive current to the test mass, a net flow of photoelectrons from the test mass to the surrounding housing surfaces. A negative apparent yield is then due to a net flow of photoelectrons from the housing to the test mass. Note that the apparent yield differs from the quantum yield (the number of photoelectrons emitted per absorbed photon, for an individual surface), and is completely dependent on the state of the system at the time of illumination. The apparent yield varied with the lamp used, applied DC biases, the AC actuation scheme and with test mass charge. With the system in a particular actuation state an entire discharge curve for an individual lamp could be measured, essentially showing how the apparent yield varied with test mass potential (charge).

\begin{figure}[]
\centering
\includegraphics[width=0.45\textwidth]{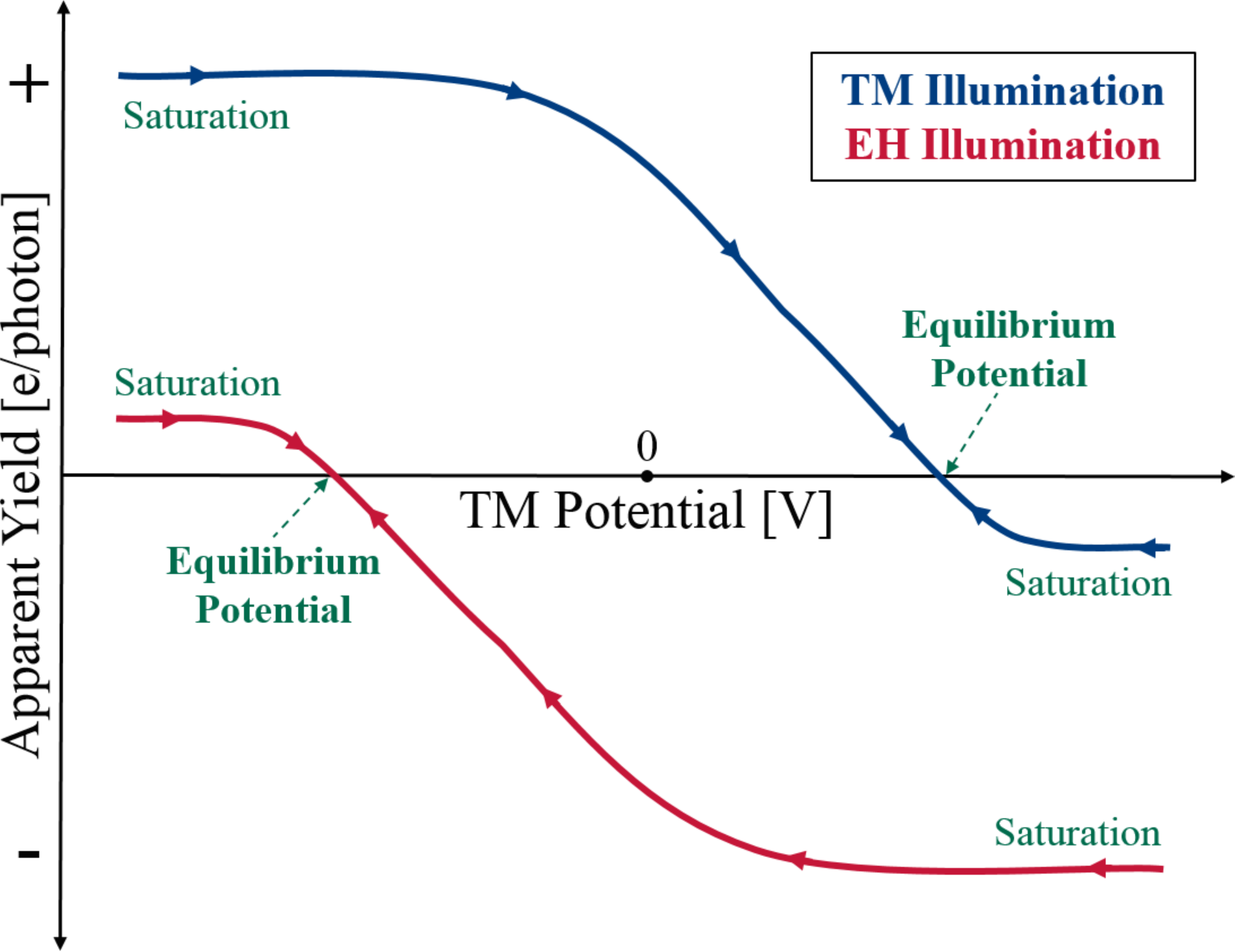}
\caption[Discharge curve cartoon.]{\label{fig:dischargeCurveCartoon} A cartoon of idealised discharge curves for both test mass (TM) and electrode housing (EH) illuminations.}
\end{figure}

Fig. \ref{fig:dischargeCurveCartoon} shows idealised discharge curves for both test mass and housing illuminations. For each illumination two saturation levels are observed. At large negative potentials all the photoelectrons originating from the test mass are able to flow away from it, while none from the housing have enough energy to overcome the potential difference and reach the test mass. At large positive potential differences the converse is true. The shape of the transition between the two saturation levels is dependent on the photoelectron energy distributions of both contributions. As a net photocurrent flows the test mass charge changes and the potential moves toward equilibrium. At this point the two opposing photocurrents balance and the apparent yield is therefore zero. We refer to this as the equilibrium potential. Finally, note that at a test mass potential of 0\,V illuminating the test mass produces a positive apparent yield (net electrons away from test mass) while illuminating the housing produces a negative apparent yield (net electrons toward test mass). Continuous discharge in science mode required a negative apparent yield (and ideally bi-polar capability) in order to keep the test mass potential around a set-point of 0\,V while countering the positive environmental charging.

During fast discharges the curves can effectively be shifted by applying DC voltages to the sensing electrodes which bias the test mass. By illuminating only when biases are temporarily applied, such a technique can aid the desired direction of discharge. It should be noted however that care needs to be taken when applying DC voltages to particular electrodes as if a significant number of photoelectrons are generated in their local region the shape of the curve can be unintentionally changed in addition to causing a shift.

In these idealised examples presented in Fig. \ref{fig:dischargeCurveCartoon}, for a particular illumination the magnitude of the saturation level is considerably higher in the intended direction of discharge than the reverse direction. In reality this is not necessarily the case. Both the geometry and reflective properties can conspire such that opposing surfaces absorb similar amounts of light leading to similar photocurrent magnitudes. Alternatively, even if the amount of light in the desired direction dominates this can be undermined if the quantum yield of the opposing surfaces are significantly higher. Such a scenario can prevent bi-polar discharge at 0\,V without the aid of DC biases.

Measurements in 2010 using the torsion pendulum facility at the University of Trento \cite{Cavalleri2009}, with a prototype GRS electrode housing and a gold coated test mass, demonstrated the difficulty of negatively charging a neutral test mass without applying additional voltages \cite{Antonucci2010}. The results were initially surprising in that bipolar discharge had been demonstrated successfully in several previous tests with similar gold coated prototype GRS hardware \cite{Wass2006, Tombolato2008}. The asymmetry of the discharge process was attributed to an imbalance in the quantum yields of the specific test mass and electrode housing surfaces under test \cite{Hollington2011}.

Prior to flight no system had been observed where fast discharge was impossible: DC biases applied to the sensing electrodes had always managed to shift the test mass potential to sufficiently suppress the unwanted photocurrent. However, application of DC biases was not desirable in science mode and this would have made a continuous discharge mode unusable. Efforts to create more reproducible properties by baking surfaces under vacuum led to mixed results \cite{Hechenblaikner2012}.

There were obvious concerns how such unpredictable variations in the quantum yield could affect the flight discharge system and a dedicated study looked at various ways of mitigating the risk. Possibilities included doping patches of the gold surface to produce higher but more predictable quantum yields or adding mirrors to the ISUK tips in order to redirect the light. However, the additional risks to the primary mission with both options were deemed too high and a more pragmatic solution was agreed. The handling and storage of sensor surfaces were strictly controlled and at final integration on ground the discharge properties were measured before and after the system underwent a low temperature bake-out under vacuum. The system then remained sealed but not actively pumped until it was vented to space when in orbit.

\section{Discharge Simulation}\label{sec:Sim}

The complexity of the problem and the desire to assess differences between ground and in-flight measurements motivated the need to simulate the system. The first full simulation was created at Imperial College London and was described in \cite{Hollington2011} with another full model of the system developed independently by Airbus Defence and Space \cite{Ziegler2014}. The updated Imperial College simulation described here consists of two parts. First, the propagation and absorption of the UV light within the sensor is modelled with a ray tracing algorithm. The ray trace output can then be fed into the second part of the simulation which approximates the sensor surfaces as a series of parallel plates. It calculates how many photoelectrons can flow given their energy distribution and the instantaneous electric fields present, before stepping in time and re-evaluating.

\subsection{Ray Trace}\label{sec:RayTrace}

The ray trace is written within the GEANT4 framework which is more commonly used to simulate the passage of high energy particles through matter using Monte-Carlo techniques \cite{Agostinelli2003, Allison2006, Allison2016}. It also includes a parametrisable micro-facet based reflection model and the ability to simulate arbitrary geometries and light sources. Using GEANT4 we model the propagation of the UV light within the sensor and calculate the percentage of total light absorbed by each distinct surface. As we shall see in the next section, we are concerned with which regions experience different instantaneous electric fields due to their effect on the emitted photoelectrons. As such we end up defining relatively large areas where we calculate the percentage of absorbed light, for example each sensing electrode, where a table of such totals forms the ray trace output.

\begin{figure}[]
\centering
\includegraphics[width=0.5\textwidth]{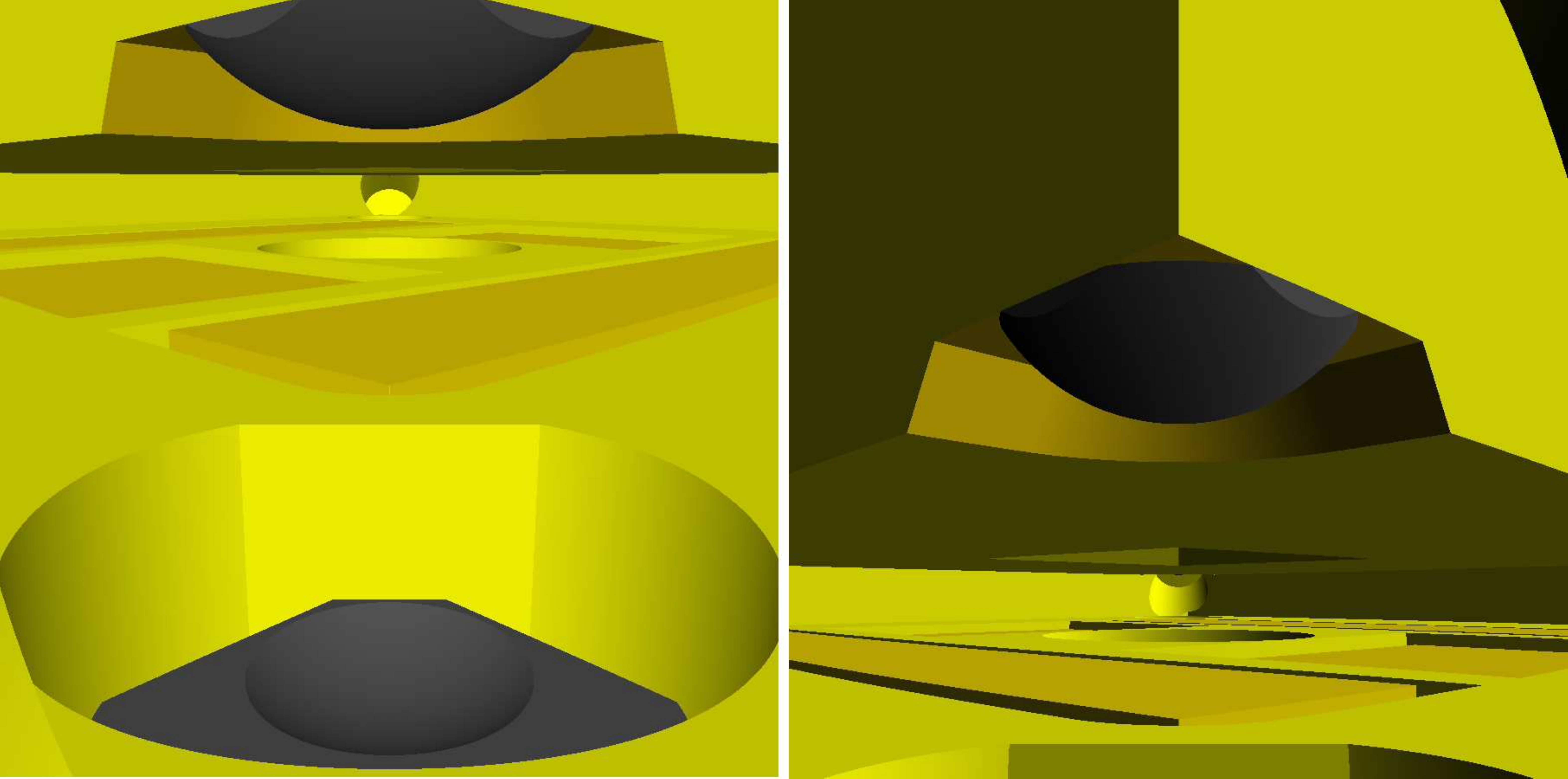}
\caption[GRS ISUK Views.]{\label{fig:ISUK_Views} Left: A view of the sensor as seen from an ISUK aimed at the housing. A retracted caging finger can be seen in the foreground with a concave hemisphere visible at its tip. With the finger engaged, this feature fit over the dome structure that can be seen just above on the corner of the test mass. Right: A view of the sensor as seen from an ISUK aimed at the test mass. The uncoated corner dome is clearly visible as well as part of the $x$, $y$ and $z$-faces of the test mass.}
\end{figure}

The model includes the full flight geometry with the option of having the central plungers either engaged or retracted. Any of the six ISUKs can be used as a light source with the emitted light cone from each having been experimentally measured. It was found that the emission cone angle was strongly dependant on the length of the spacer chosen for each optical chain causing the cone half angles to vary from $7.4^{\circ}$ to $19.2^{\circ}$. The ISUKs themselves entered the sensors in the corners of the lower $z$-face and were angled $\pm20^{\circ}$ to the $z$-plane. All results presented here used $10^{7}$ randomly generated initial rays with this number producing results repeatable to 2 decimal places for each ISUK illumination. Fig. \ref{fig:ISUK_Views} shows a view of the simulated sensor, as seen by either an ISUK aimed at the electrode housing or test mass.

The sensors contained three types of surface material; gold (the vast majority of surfaces by area), iridium (retracted corner caging fingers) and gold-platinum alloy (uncoated test mass corner domes and retracted central plungers). The refractive indices (which determine the specular reflection properties) of each material take literature values \cite{Johnson1972, Palik1998}. In the case of gold, a Pathfinder sample surface had also been measured and found to be in good agreement with literature values \cite{Hollington2011}. The surfaces within the sensor had a range of roughnesses that determine how light was scattered after undergoing reflection. These surface properties are summarised Table \ref{tab:Roughness}.

\begin{table}[]
\begin{center}
\begin{tabular}{ | c | c | c | l |}
\hline
Surface               & Material               &   $R_{a}$       & Comments         \\
\hline
Test Mass Sides       & Au                     & 0.6 to 0.8\,nm  & Measured.        \\
Test Mass Domes     	& Au$_{0.7}$Pt$_{0.3}$   & $<250\,$nm      & Estimated.       \\
Sensing Electrodes    & Au                     & 30 to 40\,nm    & Measured.        \\
Injection Electrodes  & Au                     & 50 to 80\,nm    & Measured.        \\
Electrode Housing     & Au                     & $ 20\,$nm       & Measured.       \\
Caging Fingers        & Ir                     & $<200\,$nm      & Estimated.       \\
Central Plungers      & Au$_{0.8}$Pt$_{0.2}$   & $<200\,$nm      & Estimated.       \\
\hline
\end{tabular}
\caption[Roughness of sensor surfaces.]{\label{tab:Roughness} The different surface types within the sensor. The parameter $R_{a}$ describes the surface roughness and is defined as the arithmetical mean deviation about the surface's mean height. While this parameter was measured for some flight surfaces only estimates were available for others.}
\end{center}
\end{table}

An approximate criterion for the appearance of non-specular behaviour is when the scale of surface features are comparable or greater than the wavelength of illuminating light \cite{Torrance1967}. Given the scale of the surface roughness and that the UV light had a wavelength of 254\,nm, reflection from the test mass sides, electrodes and housing surfaces were expected to be specular. Depending on the final roughness, the test mass corner domes, caging fingers and central plungers may have had a diffuse component but were still expected to be dominated by specular reflection. The ray trace has the capability of modelling a non-specular lobe component using a micro-facet model described with a Gaussian angular distribution and parameterised by $\sigma_{\alpha}$. However, without a direct measurement it is difficult to predict an appropriate $\sigma_{\alpha}$ and in any case only estimated $R_{a}$ values were available. To assess the impact of these unknowns the ray trace was run with a range of reflection properties for the rougher surfaces to determine how the absorption percentages would be affected. Even when the surfaces were described as completely diffuse reflectors (an unrealistic extreme) the final results only differed from the completely specular case at the level of a few percent. For simplicity we present here the model results that consider all surfaces as completely specular, but this approximation should have little affect on the final results. Likewise, the effect of uncertainties in the refractive indices, sensor geometry and ISUK output distributions are difficult to accurately quantify although adjusting each by feasible amounts has shown that none change the results significantly.

\begin{figure}[]
\centering
\includegraphics[width=0.5\textwidth]{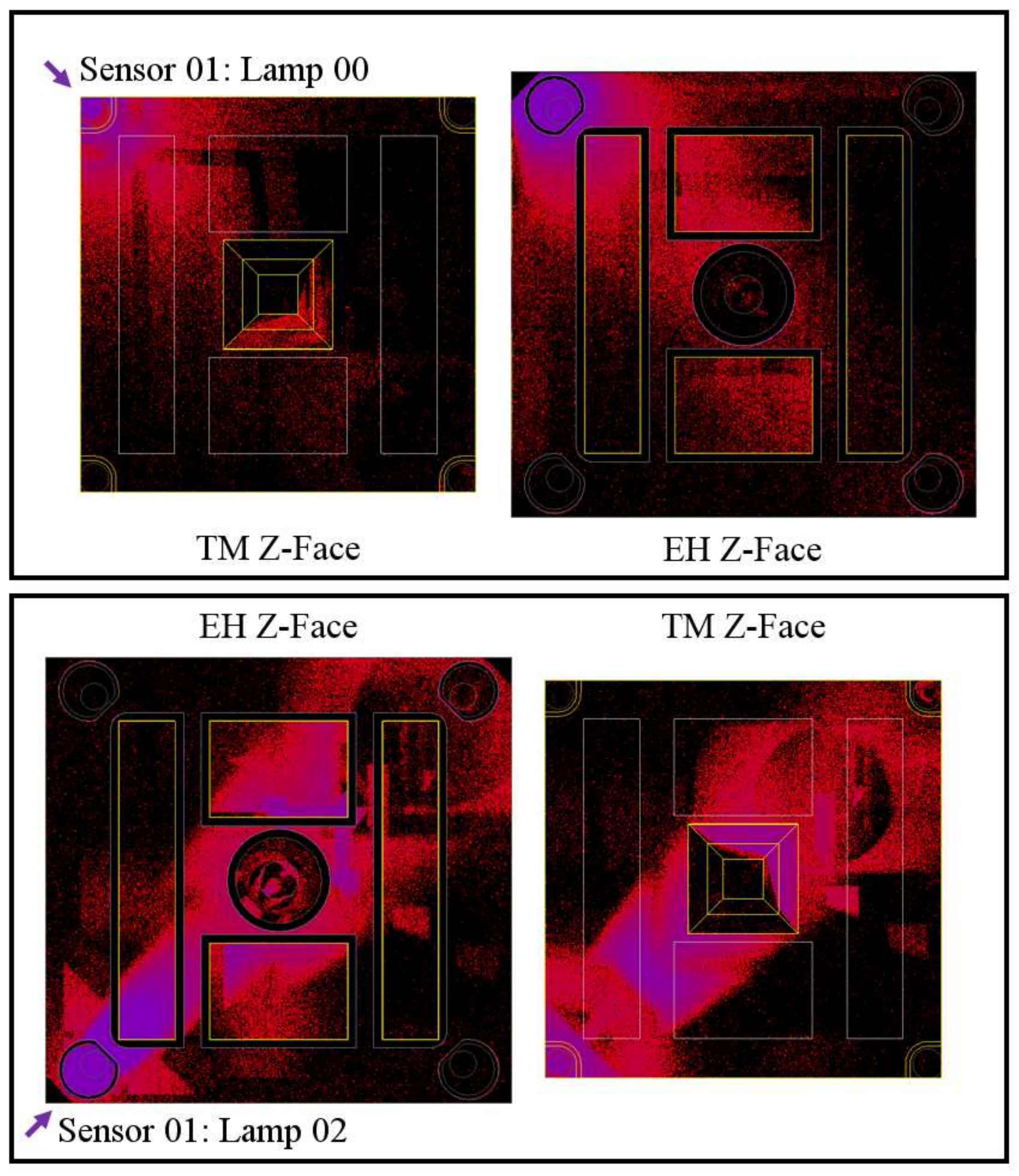}
\caption[Light distributions.]{\label{fig:lightDistr} The distribution of absorbed UV light on the lower $z$-faces of the flight sensor. Top: Illumination provided by lamp 00 which is aimed at the test mass. The primary test mass surfaces are shown on the left while light reflected onto the housing surfaces can be seen on the right. Note that the primary light falls directly on the test mass corner with little reflected light reaching the electrode regions. Bottom: Illumination provided by lamp 02 which is aimed at the housing. The primary housing surfaces are shown on the left while light reflected onto the test mass surfaces can be seen on the right. Note that the primary light is mainly absorbed within the recess of the caging finger but some light also reaches the rest of the $z$-faces.}
\end{figure}

The output of the ray trace using the flight geometry and all six ISUK illuminations are shown in Appendix \ref{tab:UVabsorption} with a colour-map representation of the data for lamps 00 and 02 (which pointed at the test mass and electrode housing respectively) shown in Fig. \ref{fig:lightDistr}. The table shows that the majority of the light is absorbed by structures on the negative $z$-faces of the test mass and housing and there is effectively no absorption by positive $z$-electrodes for any illumination. For both test mass illuminations there is also a significant amount of light (24\% and 36\%) absorbed on the $x$ and $y$-faces. Notice also the difference in results for ISUKs pointing in the same direction which is predominantly caused by each ISUK having a different light emission cone. What is most striking is that upon direct illumination of the housing around 60\% of the total UV light is absorbed within gaps between the electrode and surrounding guard ring surfaces, most of which is within the recess that the caging finger closest to illumination is retracted into. A further 11-12\% is also absorbed by the iridium caging finger itself. As will be discussed in the next section, this absorbed UV light is effectively wasted when it comes to discharging. Such sub-optimal illumination is due to the original design of the caging mechanism (which did not include the corner caging fingers) being reviewed after the ISUK entry points had been finalised. This has also had consequences for the test mass illumination where the corner domes absorb a significant amount of light (around 43\% and 23\%). These general findings can be seen visually within the colour-maps.

\subsection{Photoelectron Flow Model}\label{sec:PEflow}

Written in MATLAB, the photoelectron flow model uses the output of the ray trace to estimate the apparent yield of the system, given the photoelectric properties of individual surfaces as well as any applied AC or DC electric fields. The model simplifies the problem by splitting the sensor up into pairs of adjacent surfaces on the test mass and electrode housing and treating them like parallel plates with a uniform electric field between them. The model calculates the net photoelectron flow between each surface pair while taking into account any instantaneous electrical potential difference as well as the energy distribution of the emitted photoelectrons. The system is then stepped in time, applied AC voltages recalculated and the test mass potential updated due to any charge transfer in the previous step. The net flow is then evaluated again with this process repeated for a user defined number of steps.

Though photoelectrons are emitted over a range of angles (described by a cosine distribution), the model only considers the component of an electron's kinetic energy that is normal to the emitting surface. As shown in \cite{Hechenblaikner2012}, this is well described by a triangular distribution with an $E_{max}$ equal to the difference between the UV photon energy (4.89\,eV) and the surface work function, with the distribution peaking at around $E_{max}/5$. As the model approximates the surfaces as a series of parallel plates it is this energy component that will determine if a photoelectron can overcome any opposing electric potential. Such a triangular distribution has also been confirmed by unpublished measurements made at Imperial College London. 

The model includes the 100\,kHz, 4.88\,V voltage applied to the six injection electrodes as well as the translational and rotational actuation voltages applied to the twelve sensing electrodes at frequencies of 60\,Hz, 90\,Hz, 120\,Hz, 180\,Hz, 240\,Hz and 270\,Hz \cite{Weber2003}. In-flight the amplitude of these voltages varied depending on the axis and particular actuation mode but were typically around 2\,V during the charge measurements that will be presented. All actuation biases were applied to pairs of electrodes with opposite polarity so as not to bias the test mass whereas the injection bias was applied with the same polarity to all electrodes causing the test mass potential to oscillate by $\pm0.6$\,V at 100\,kHz. Given typical photoelectron energies of order 0.1\,eV (equivalent velocity $\sim2\times10^{5}\,\textup{ms}^{-1}$) and the largest gap being 4\,mm, crossing times should be of order 20\,ns. The highest frequency AC voltage was 100\,kHz, with a period that is therefore equivalent to 500 crossing times such that the AC fields appear quasi-static to the photoelectrons. Some charge measurements included additional DC voltages applied to particular sensing electrodes to intentionally shift the test mass potential which the model also includes.

As shown by the ray trace results, a significant fraction of light is absorbed within housing gaps and recesses, particularly for lamps aimed at the electrode housing. Any photoelectrons generated from these surfaces have no adjacent region on the test mass and are instead opposite either another housing surface or the iridium caging finger. For two reasons it is very unlikely that these photoelectrons will reach the test mass. Firstly they are emitted randomly with a cosine distribution meaning that at most only half will be directed toward the test mass. From the photoelectron's perspective, the deeper within a gap it originates the smaller the solid angle the test mass subtends making it even less likely. The second and more significant reason results from work function differences between the iridium caging fingers and the surrounding gold coated recess and to a lesser extent the small differences in work function across the gold surfaces within the gaps (so called patch potentials \cite{Weber2007}). These differences lead to contact potentials (also known as Volta potentials) of possibly a few tenths of a volt between gold surfaces and possibly up to a volt between the gold and iridium \cite{Speake1996}. Given photoelectron energies of less than 1\,eV from gold, electric fields within the gaps resulting from these contact potentials would have a strong influence on behaviour. The upshot of these effects is that it is very unlikely that light absorbed within housing gaps can produce photoelectrons that can reach the test mass. Therefore the model simply ignores any UV light absorbed within the gaps as these can not affect the net flow of charge to and from the test mass.

As mentioned, the model simplifies the electric field geometries within the sensor by only considering parallel field lines between adjacent surfaces. While uniform fields are an excellent approximation at the centre of the electrode and test mass faces, toward the edges the field geometries become increasingly complicated. This is particularly true around the caging features and where the ray trace has already shown the majority of the light is absorbed. Fortunately, it is a photoelectrons final destination that is important rather than its exact trajectory and given their relatively low energy compared to the typical field strengths this tends to be inevitable. Treating the fields in this way significantly reduces the simulation run time, vital when fitting in a large parameter space. Similarly, a small amount of light (less than a few of percent) is absorbed by surfaces at the edges and corner of the housing that have no directly opposite region on the test mass. In these cases the first electric field the emitted photoelectrons encounter will be that between the test mass and grounded housing. Again, given their relatively low energies most of the time the photoelectrons destination will be determined simply by the direction of the field. As such UV absorption in these regions is treated as if it were simply facing the test mass.

\section{Pre-flight Measurements on Ground}\label{sec:Gmeasure}

Measurements of the flight sensor's discharge properties were carried out on ground between September and October 2014 by OHB Italia SpA (formally Compagnia Generale per lo Spazio) in Italy. The sensors were mounted in their flight vacuum enclosures but had yet to be integrated onto the spacecraft. To allow measurements to be made the central plungers were engaged and used to hold the test masses but the eight caging fingers were retracted so as to affect illumination as little as possible. An electrometer (Keithley, model 6517B) was connected to the otherwise isolated plungers and used to directly measure the drain current from the test mass during illumination, unlike in-flight where changes in the test mass potential with UV illumination were the only measurement possible.

{ % Shaded cells have gaps at edges. As a workaround, reduce column widths just for this table by placing command within curly brackets.
\setlength{\tabcolsep}{0pt}
\begin{table*}[]
\begin{center}
\begin{tabular}{c | c | c | c | c |}
\cline{2-5}
                                & \multicolumn{4}{c|}{Apparent Yield}	                                     						    					    \\
	 			      & \multicolumn{4}{c|}{(net electrons/injected photon)}	         				                       					\\
\cline{2-5}
       & \multicolumn{2}{c|}{Before Bake-out}                 & \multicolumn{2}{c|}{After Bake-out}            \\
\cline{2-5}
                        & $-5$\,V & $+5$\,V & $-5$\,V & $+5$\,V \\
			     & TM Saturation & EH Saturation & TM Saturation & EH Saturation \\
\hline
\multicolumn{1}{| c |}{~Sensor 01: Lamp 00 (TM)} & \cellcolor{Gray} $(+1.67\pm0.03)\times10^{-5}$ & $(-1.03\pm0.05)\times10^{-6}$                  & \cellcolor{Gray} $(+2.49\pm0.08)\times10^{-5}$ & $(-1.91\pm0.06)\times10^{-6}$                  \\
\multicolumn{1}{| c |}{~Sensor 02: Lamp 01 (TM)} & \cellcolor{Gray} $(+4.65\pm0.54)\times10^{-5}$ & $(-1.35\pm0.32)\times10^{-6}$                  & \cellcolor{Gray} $(+2.18\pm0.03)\times10^{-5}$ & $(-1.99\pm0.05)\times10^{-6}$                  \\
\hline
\multicolumn{1}{| c |}{~Sensor 01: Lamp 02 (EH)} & $(+2.52\pm0.03)\times10^{-6}$                  & \cellcolor{Gray} $(-3.34\pm0.12)\times10^{-6}$ & $(+2.47\pm0.03)\times10^{-6}$                  & \cellcolor{Gray} $(-4.19\pm0.03)\times10^{-6}$ \\
\multicolumn{1}{| c |}{~Sensor 01: Lamp 04 (EH)} & $(+2.35\pm0.05)\times10^{-6}$                  & \cellcolor{Gray} $(-1.87\pm0.03)\times10^{-6}$ & $(+2.84\pm0.05)\times10^{-6}$                  & \cellcolor{Gray} $(-3.67\pm0.05)\times10^{-6}$ \\
\multicolumn{1}{| c |}{~Sensor 02: Lamp 05 (EH)} & $(+6.06\pm0.96)\times10^{-6}$                  & \cellcolor{Gray} $(-2.48\pm0.54)\times10^{-6}$ & $(+2.16\pm0.03)\times10^{-6}$                  & \cellcolor{Gray} $(-2.33\pm0.04)\times10^{-6}$ \\
\multicolumn{1}{| c |}{~Sensor 02: Lamp 06 (EH)} & $(+7.90\pm1.24)\times10^{-6}$                  & \cellcolor{Gray} $(-4.28\pm0.75)\times10^{-6}$ & $(+2.82\pm0.11)\times10^{-6}$                  & \cellcolor{Gray} $(-5.12\pm0.20)\times10^{-6}$ \\
\hline
\hline
\multicolumn{1}{| c |}{CoV (TM)}                & \cellcolor{Gray} $66.7\%$                      & $19.0\%$                                       & \cellcolor{Gray} $9.4\%$                       & $2.7\%$                                        \\
\hline
\multicolumn{1}{| c |}{CoV (EH)}                & $58.0\%$                                       & \cellcolor{Gray} $35.1\%$                      & $12.6\%$                                       & \cellcolor{Gray} $30.4\%$                      \\
\hline
\end{tabular}
\caption[Flight AYs on ground.]{\label{tab:flightAYground} The flight sensor apparent yields measured before and after bake-out, with statistical uncertainties. Each of the six lamps were directed at either the test mass (TM) or electrode housing (EH) but inevitably generated a photocurrent in both directions. By biasing the test mass by either $\pm5$\,V the aim was to completely suppress one of these, saturating the net photocurrent in the opposite direction. A positive apparent yield represents a saturated photocurrent from the test mass while a negative apparent yield represents a saturated photocurrent from the housing. The shaded cells denote the saturated apparent yield with direct illumination. Note that the apparent yields for direct test mass illumination are an order of magnitude higher than the other measurements. The coefficient of variation $(\sigma/\mu)$ was also calculated (ignoring statistical uncertainties) for appropriate groups of measurement to quantify how the surface properties varied for a given illumination.}
\end{center}
\end{table*}
}

To perform the measurements each of the two sensor vacuum enclosures were in turn mounted in a test-rig thermal-vacuum chamber. Illumination was provided by a single mercury lamp and an optics barrel identical to that used in the ULU but two off-the-shelf fibres and a vacuum feed-through were used to route the light to each ISUK. A mechanical chopper was placed in the optical chain to modulate the light signal and allow for a lock-in measurement that improved sensitivity. Unlike in-flight, the light cone emitted from each ISUK was nominally identical and a series of calibration measurements with a UV power meter were used to determine the throughput of the optical chain. During the test all electrodes were electrically shorted to the housing with the isolated test mass in electrical contact only with the plungers. For each ISUK illumination a measurement was made with the test mass biased at both $-5$\,V and +5\,V with respect to the housing. With photoelectron energies of around 1\,eV these biases are sufficient to completely turn-off one of the two opposing photocurrents, allowing extraction of the saturation levels illustrated in Fig. \ref{fig:dischargeCurveCartoon}. These measurements were repeated before and after a 24 hour, $115\,^{\circ}\textup{C}$ bake-out under vacuum (with a further 24 hours either side for heating and cooling of the system). The results are shown in Table \ref{tab:flightAYground}.

After first distinguishing between before and after bake-out, the measurements can be gathered into four groups; the saturated apparent yield of the test mass surfaces with direct illumination (one per sensor), the saturated apparent yield of the test mass surfaces with indirect illumination (two per sensor), the saturated apparent yield of the electrode housing surfaces with direct illumination (two per sensor) and the saturated apparent yield of the electrode housing surfaces with indirect illumination (one per sensor). Within each of these groups the distribution of absorbed UV light was nominally identical, therefore any variation within a particular group should only be due to differences in the photoelectric properties of each illuminated area. Comparing these groups before and after the bake-out shows that there was no clear systematic increase in the apparent yields with some rising and some falling. However, what did occur was a reduced variation in the apparent yields within each group; not only between different corners of a single sensor but even between the two separate sensors. This is confirmed if the coefficient of variation (CoV) is considered for each group (the ratio of standard deviation and mean value, provided here as a percentage), with three out of the four groups showing a significant reduction after bake-out. The one exception is for the electrode housing when directly illuminated, which showed a comparable level of variation before ($35.1\%$) and after bake-out ($30.4\%$). The general increase in uniformity was presumably due to a change in the surface state; either the heating led to a more uniform redistribution of surface contaminates or it removed weakly bound adsorbates with the remaining tightly bound layers common to all surfaces. Finally, considering just the results after bake-out the apparent yield for the test mass surfaces with direct illumination were around a factor 6 greater than when the electrode housing surfaces were directly illuminated. As we shall see in Sec. \ref{sec:PEproperties}, to explain these differences in apparent yield the ray trace is required to disentangle the affect of the total absorbed light from the intrinsic quantum yields.

It should be noted that there were several systematic differences between the measurements made on ground and the situation in-flight. First, the optical path between the lamp and the ISUKs differed from that in-flight producing a light emission cone with a half angle of $12.3^{\circ}$, common to all the ISUKs. This change meant slightly different regions within the sensor were inevitably illuminated, with potentially different photoelectric properties. The second difference was that during flight the central plunger was retracted and was electrically connected to the housing rather than the test mass. This had three consequences; it changed the light distribution within the sensor slightly, altered the electric field distribution in its vicinity and any photoelectrons generated from it acted to charge the test mass negatively. The final difference, and arguably most significant, was the absence of the various time varying sensing and actuation voltages present within the flight sensors. Experience from the Trento tests on GRS prototype hardware showed that they have a significant influence on the overall discharging behaviour \cite{Hollington2011}. By using the models described in Sec. \ref{sec:Sim}, which can account for these differences, the ground and flight tests can however be analysed together and their intrinsic surface quantum yield values compared. The results of this comparison will presented in Sec. \ref{sec:PEproperties}.

\section{In-flight Measurements}

During ground testing the entire system was pumped down to $<10^{-5}\,\textup{mbar}$ and the design of the flight caging and venting mechanism allowed each vacuum enclosure to be closed while the outer chamber of the test rig was still under vacuum. The sensors were then left sealed (but with no active pumping) throughout spacecraft integration with the internal pressure estimated to have drifted up to around a millibar during this time. Following a successful launch on $3^{\textup{rd}}$ December 2015, Pathfinder reached its operational orbit around the L1 Lagrange point in February 2016. The vacuum chambers housing the sensors were then vented to space on $3^{\textup{rd}}$ February, over 15 months after the discharge measurements were made on ground. 

A series of initial calibration measurements showed that despite the systematic differences compared to the ground test, the discharge properties were qualitatively similar to those measured pre-flight. As before lamps 00 and 01, which were aimed at the test masses, had significantly higher apparent yields than the lamps aimed at the electrode housing. As will be described in Sec. \ref{sec:PEproperties}, dedicated measurements made later in the mission quantified these differences and allowed comparisons to be made with simulation. During these initial measurements it was also discovered that during spacecraft integration the fibre chains for lamps 01 and 05 had been inadvertently switched. Thus lamp 05 was unintentionally attenuated by a factor of $\sim63$, while lamp 01 had no attenuation. Nevertheless, both were still capable of discharging although lamp 05 at a significantly reduced rate.

Following calibration, the CMS was commissioned for the start of the mission proper on the $1^{\textup{st}}$ March 2016 and operated successfully until the mission's end on the $17^{\textup{th}}$ July 2017. The six lamps were turned on a combined total of 418 times in-flight (without fail) and were operated for a total of 421 hours. No significant signs of ageing, such as output instability, were observed for any of the lamps. However, by the end of the mission usage levels were only just approaching those where such effects were predicted to start.

After initial release from the caging mechanism, residual test mass charges equivalent to $-421$\,mV and $-281$\,mV was measured on the test masses. The level of residual charge is related to the difference in work function of the last two surfaces in electrical contact as well as triboelectric processes. For reasons that remain unclear, subsequent grabs and releases throughout the rest of the mission produced residual test mass potentials systematically lower and ranged from $-98$\,mV to $-52$\,mV. During typical science mode operations the charging rate for both test masses was similar and drifted upward from around $+21\,\textup{es}^{-1}$ at the start of the mission to around $+36\,\textup{es}^{-1}$ by the end of the mission. This increase was predominately due to the declining solar cycle resulting in a gradually increasing GCR flux. Such rates were in good general agreement with pre-flight predictions \cite{Wass2005} but dedicated experiments revealed additional charging rate dependencies on actuation scheme and test mass potential, with rates as high as $+57\,\textup{es}^{-1}$ measured at an extreme test mass potential of $-1$\,V. Understanding these dependencies will form the basis of a future paper but suffice to say that charging typically caused the test mass potential to increase by around $+12$\,mV per day.

For the vast majority of the mission a fast discharge scheme was employed. During a weekly spacecraft station keeping period the CMS was used to discharge the test masses to a set level, for example $-80$\,mV, taking around 30 minutes. Incident cosmic rays then gradually charged the test masses positively until the next station keeping; at which point the CMS reset the test masses' charge to the desired level. At the start of the mission this was performed weekly but was later reduced to every two or three weeks allowing longer uninterrupted science runs. Initially these fast discharges were done open-loop by bringing the test mass to an equilibrium potential that had been shifted by applying appropriate DC voltages during the discharge. While this method was both simple and reliable it was also fairly inefficient when it came to lamp usage. Once confidence in the system's performance had increased these routine fast discharges were instead handled by the automated closed-loop CMS software. Both methods typically got within the equivalent of a few millivolts of the desired test mass charge level, similar in scale to the systematic uncertainty in the measurement due to stray DC voltages on the sensing electrodes.

Two experiments were also performed to test the continuous discharge concept, successfully holding both test mass potentials within 10\,mV of zero for several days. A detailed analysis of these tests will be presented in a future article.

\subsection{Long Term Stability}

As explained previously, the apparent yield for a particular lamp is dependent on the state of the system at the time of the measurement. To make a clean assessment of the stability of the photoelectric properties over course of the mission a sub-set of measurements need to be chosen that were made with the system in a reproducible state. The automated discharges carried out during the regular station keeping periods offer an ideal dataset for this. Functionally, the automated CMS software performed a discharge in the same way each time. After first measuring the test mass charge by applying a dither voltage it then calculated the appropriate lamp setting and duration to reach the requested level of final test mass charge. Following illumination it would remeasure the test mass charge and determine if it was within a predefined range of success. If the system had over/under shot the lamp settings would be recalculated and the sensor illuminated for a second time. This would continue until the desired discharge level had been reached or the time allocated for discharging had been exhausted. To aid the discharge, the CMS applied relatively large DC voltages to the sensing electrodes ($\pm5\,\textup{V}$ on $x$ and $y$, $\pm1\,\textup{V}$ on $z$) that shifted the test mass potential by approximately $\pm1.27\,V$.

\begin{figure}[]
\centering
\includegraphics[width=0.5\textwidth]{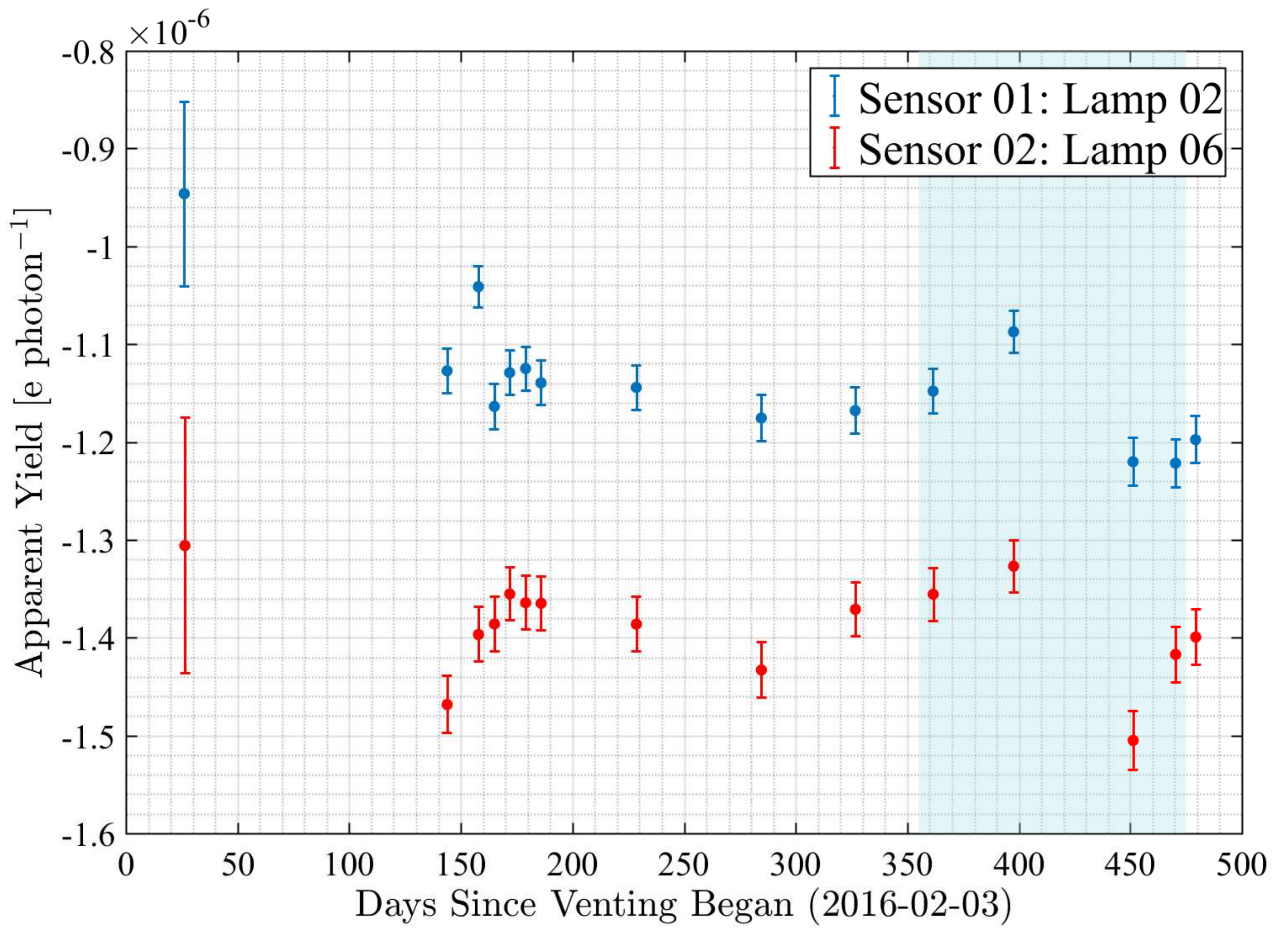}
\caption[Discharge Stability.]{\label{fig:dischargeStability} The apparent yields observed during CMS fast discharges for lamp 02 coupled to sensor 01 and lamp 06 coupled to sensor 02. The first two measurements were made during commissioning 26 days after venting began. They contain a greater level of uncertainty but allow the time baseline to be extended significantly. The shaded blue regions show a four month period where the sensor temperatures were lowered to $12\pm1\,^{\circ}\textup{C}$ due to an unrelated experiment. For the majority of the mission the sensors were at $23\pm1\,^{\circ}\textup{C}$.}
\end{figure}

By June 2016 the automated CMS was in routine use, typically discharging from a few positive tens of milli-volts to either $-80$\,mV or $-120$\,mV. Given the systematic nature of the discharges the CMS would typically use a particular lamp aimed at the electrode housing, lamp 02 for sensor 01 and lamp 06 for sensor 02, and also a similar lamp output setting each time. Combined with on ground calibration measurements, the telemetered photodiode readings within the lamp blocks allow the number of photons entering the sensor during a discharge to be estimated. This can then be used with the change in test mass charge measured before and after illumination to calculate the apparent yield. In addition, a couple of early in-flight test measurements were made during commissioning at the end of February 2016 which allow the baseline of long term stability to be extended. While these measurements were similar to the automated CMS discharges as a precaution the lamps were commanded manually and only on for a relatively short amount of time at a low setting. They were also made when the absolute test mass charges were lower than the other measurements considered at levels equivalent to $-275$\,mV and $-110$\,mV respectively. These two measurements therefore have a larger statistical error but also the possibility of a systematic error due to the lamp setting used and the test mass potential at the time of illumination. The apparent yields obtained during CMS fast discharges for lamps 2 and 6 are shown in Figure \ref{fig:dischargeStability}.

The apparent yield for both lamps were stable at about the $10\%$ level approximately 140 days after the system was vented to space and remained so for the duration of the mission. In both cases the apparent yields were consistent with the measurements made during commissioning, although the aforementioned caveats should be kept in mind. The small drift observed in the measurements was correlated with the test mass potential at the start of the discharge which was not unexpected. Given the systematic nature of the discharges and the fact the test masses tended to charge up in tandem, this also explains the ostensible correlation between the apparent yields for both lamps. With the DC biases applied during the fast discharges ($+5\,\textup{V}$ on $x$ and $y$, $+1\,\textup{V}$ on $z$), the measured apparent yields should have been approaching the negative saturation levels for each lamp. As this level is dominated by the quantum yield of the electrode housing we can say that it was stable with time under vacuum, cosmic ray irradiation and a temperature reduction of $11\,^{\circ}\textup{C}$ as well as with the UV irradiation from the discharging itself. It also shows that the optical chains between each lamp and the sensors did not degrade as this would have been observable in these measurements. While these measurements alone do not necessarily prove that the photoelectric properties were unchanged compared to those made on ground, they do demonstrate they reached a stable level once in space.

\subsection{Discharge Curves}\label{sec:dischargeCurves}

Although the CMS was able to successfully discharge both test masses from day one of the mission, it is vital that it is understood why the system behaved the way it did such that its success can be replicated for LISA. This was particularly important given the variability in photoelectric properties that had been measured during Pathfinder development and the fact that the sensor's geometrically complicated corner regions absorbed significant amounts of light. Several dedicated experiments were performed to characterise each lamp's full discharge curve, such as that shown in Figure \ref{fig:dischargeCurveCartoon}, which shows the change in apparent yield with test mass potential. Again, a lamp's discharge curve is dependent on the area of the sensor illuminated, the quantum yield and work function of surfaces where UV light is absorbed and the voltages present in regions where photo-emission occurs. By measuring each lamp's full discharge curve a deeper understanding of the overall system is gained, it aids modelling and allows comparison with measurements made on ground.

\begin{figure}[]
\centering
\includegraphics[width=0.49\textwidth]{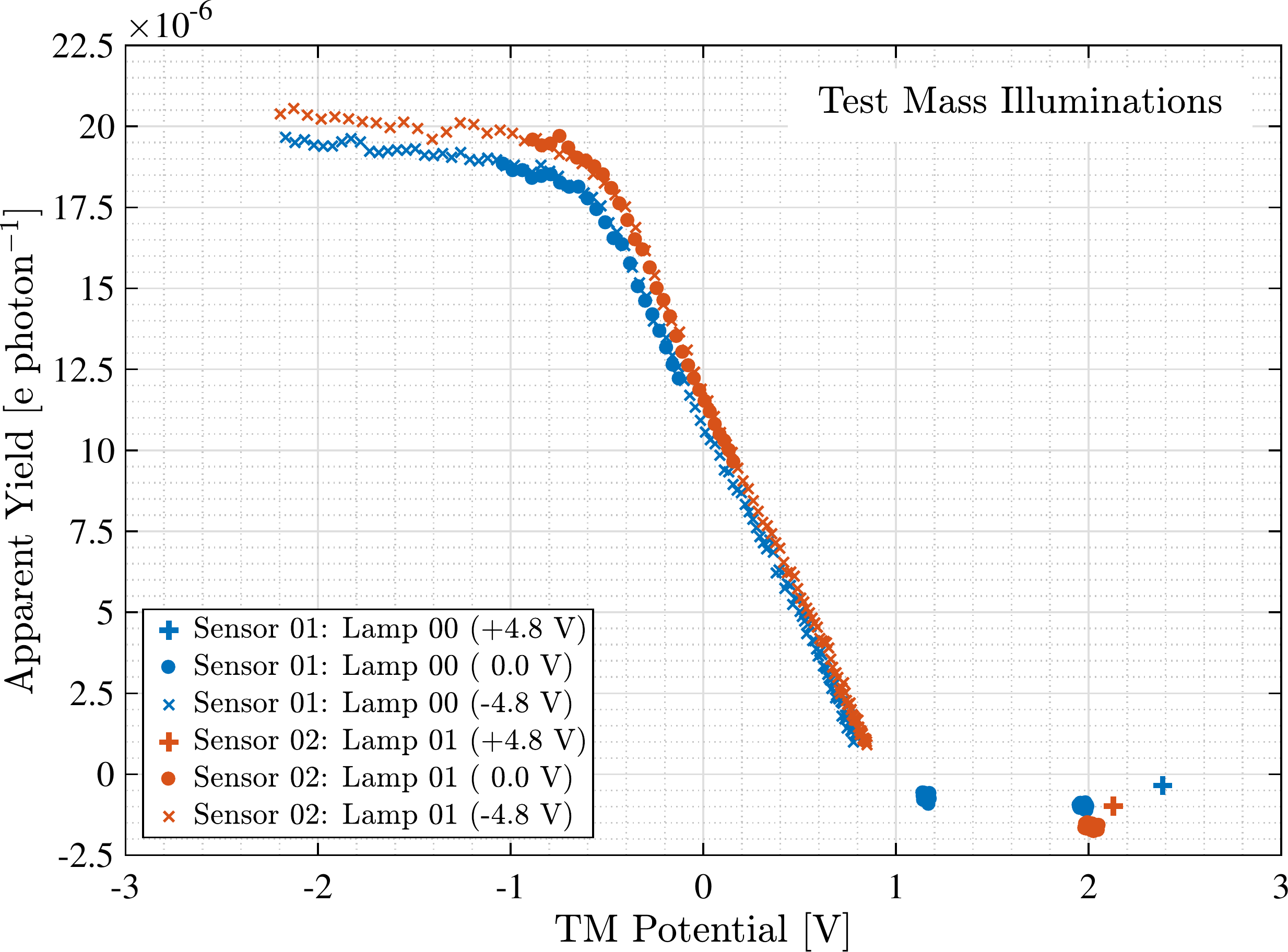}
\includegraphics[width=0.49\textwidth]{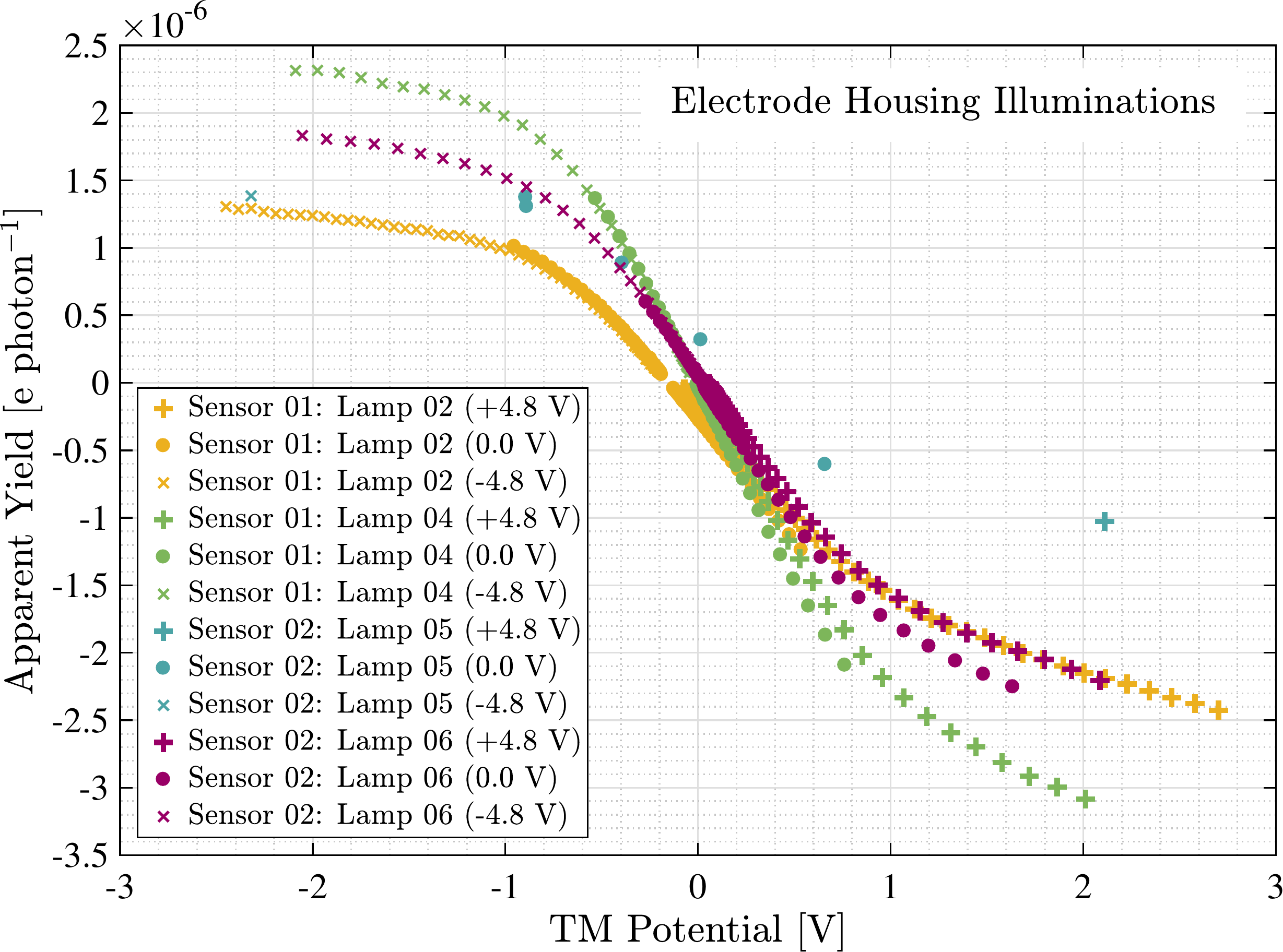}
\caption[Measured discharge curves.]{\label{fig:measuredCurves} Top: Measured discharge curves for lamps 00 and 01, which primarily illuminated the test masses in sensors 01 and 02. Bottom: Measured discharge curves for the four lamps which were aimed at the electrode housing of either sensor 01 or 02. Filled circles indicate measurements made with no applied DC biases, $+$ symbols where $+4.8\,$V was applied to all $x$ and $y$ sensing electrodes and $\times$ symbols where $-4.8\,$V was applied instead. Combined with its relatively low yield, the inadvertent attenuation of lamp 05's output made only spot measurements practical due to time constraints. Note that while the random error on individual points is small (error bars removed to improve clarity) there is a systematic uncertainty specific to each curve at the 10-20\% level. This would scale an entire curve and is due to systematic uncertainties in the calibration of the optical chains.}
\end{figure}

The measurements described here were made on the $1^{\textup{st}}$ and $2^{\textup{nd}}$ February 2017 and the $23^{\textup{rd}}$ and $24^{\textup{th}}$ June 2017. They consist of spot measurements where the test mass charge was measured before and after relatively short illuminations as well as continuous measurements made during longer illuminations. For a particular test mass, lamps were used in sequence, with one used to charge the test mass positively while the next would charge negatively. To allow higher test mass potentials to be reached, during some illuminations DC biases of $\pm4.8\,$V were applied to all $x$ and $y$ sensing electrodes that shifted the test mass potential by 1.18\,V. Strictly speaking these measurements describe a separate discharge curve to those made with no applied DC biases as they change the local electric fields around the electrodes. However, they are presented together here as shifts in the curves relate to the fraction of photoelectrons emitted in the $x$ and $y$ sensing electrode regions and allow a more stringent test of our simulations. All the measurements presented here were made with the AC actuation voltages as low as possible, with amplitudes typically less than 2\,V. Note that the long term measurements discussed in the previous section were made with significantly larger actuation voltages present leading to an expected difference in the observed apparent yields. Finally, to limit systematic differences related to a lamp's photodiode calibration all measurements made with a particular lamp were made at the same output setting. The measured discharge curves for all six lamps are shown in Figure \ref{fig:measuredCurves}.

What is immediately striking is the similarity between the curves for lamps 00 and 01, which were aimed at test mass 01 and 02 respectively. The positive part of the curve dominated by emission from test mass surfaces differ by only a few percent, although there are subtle but significant differences if the negative apparent yield from housing surfaces is considered. Nonetheless, such quantitative similarities are slightly surprising given that the optical chain for lamp 00 featured heavy attenuation by design whereas the chain for lamp 01 inadvertently had none. On its own this should not affect the discharge curves as the net change in test mass charge should scale with the number of injected photons. However, as mentioned previously the length of the attenuating spacer in each lamp's optical chain changed the angular distribution of light injected into the sensor, the greater the level of attenuation the narrower the beam emitted. Indeed, this is seen in the UV ray trace results (detailed in Sec. \ref{sec:RayTrace} and tabulated in Appendix \ref{tab:UVabsorption}) where the total absorbed by each test mass is similar (66.07\% and 64.50\% respectively) but the amount absorbed by the gold-platinum corner domes is significantly different (42.58\% and 23.17\% respectively) due to the difference in ISUK light distributions. Despite such a significant difference in where the light was absorbed the similarity in the apparent yields suggests that the uncoated gold-platinum corner domes had a similar quantum yield to that of the gold. For both lamps the equilibrium potential was just below $+1\,$V. The negative saturation yield for lamp 01 was about twice that of lamp 00 and spot measurements for both showed the level was roughly halved when $+4.8\,$V was applied compared to those with no DC bias on the $x$ and $y$ sensing electrodes. The measurements also show that the positive yield had not completely saturated by $-2.2\,$V and still showed a small negative slope.

The curves for the four lamps aimed at the electrode housing are qualitatively similar to each other but significantly different from those aimed at the test mass. For the four housing lamps, the amplitude of the negative saturation levels were comparable to those of the positive saturation levels. Partly in consequence of this, the equilibrium potential of lamps 04 and 06 were close to zero and only lamp 02 had a significantly negative equilibrium potential ($-150\,$mV) with no applied DC bias making it the only lamp suitable for the automated CMS continuous discharge. The heavily attenuated lamp 05 had a light output distribution significantly different to the other three lamps and the most positive equilibrium potential ($+250\,$mV). All four lamps had a fairly wide range in their saturation levels, although with the exception of lamp 05's positive level, none appeared to have fully saturated by $\pm2.5\,$V. Lamps 02, 04 and 06 showed clear shifts in negative apparent yields when $+4.8\,$V was applied compared to those with no DC bias on the $x$ and $y$ sensing electrodes.

Comparing the discharge curves to the idealised cases shown in Figure \ref{fig:dischargeCurveCartoon}, both lamps primarily illuminating the test mass produced curves in qualitative agreement with expectation. The positive saturation level had an apparent yield over $10\times$ higher than that at the negative level leading to a positive apparent yield at a test mass potential of 0\,V. However, none of the four lamps aimed at the electrode housing matched the ideal case. While the positive saturation levels were comparable in amplitude to the negative levels observed with test mass illumination, the negative levels were around an order of magnitude lower than the equivalent obtained with test mass illumination. While this behaviour was not unexpected based on the ground tests it was non-optimal and to fully understand it the simulated system described previously needs to be invoked.

\subsection{Estimated Photoelectric Properties}\label{sec:PEproperties}

\begin{figure*}[]
\centering
\includegraphics[width=0.49\textwidth]{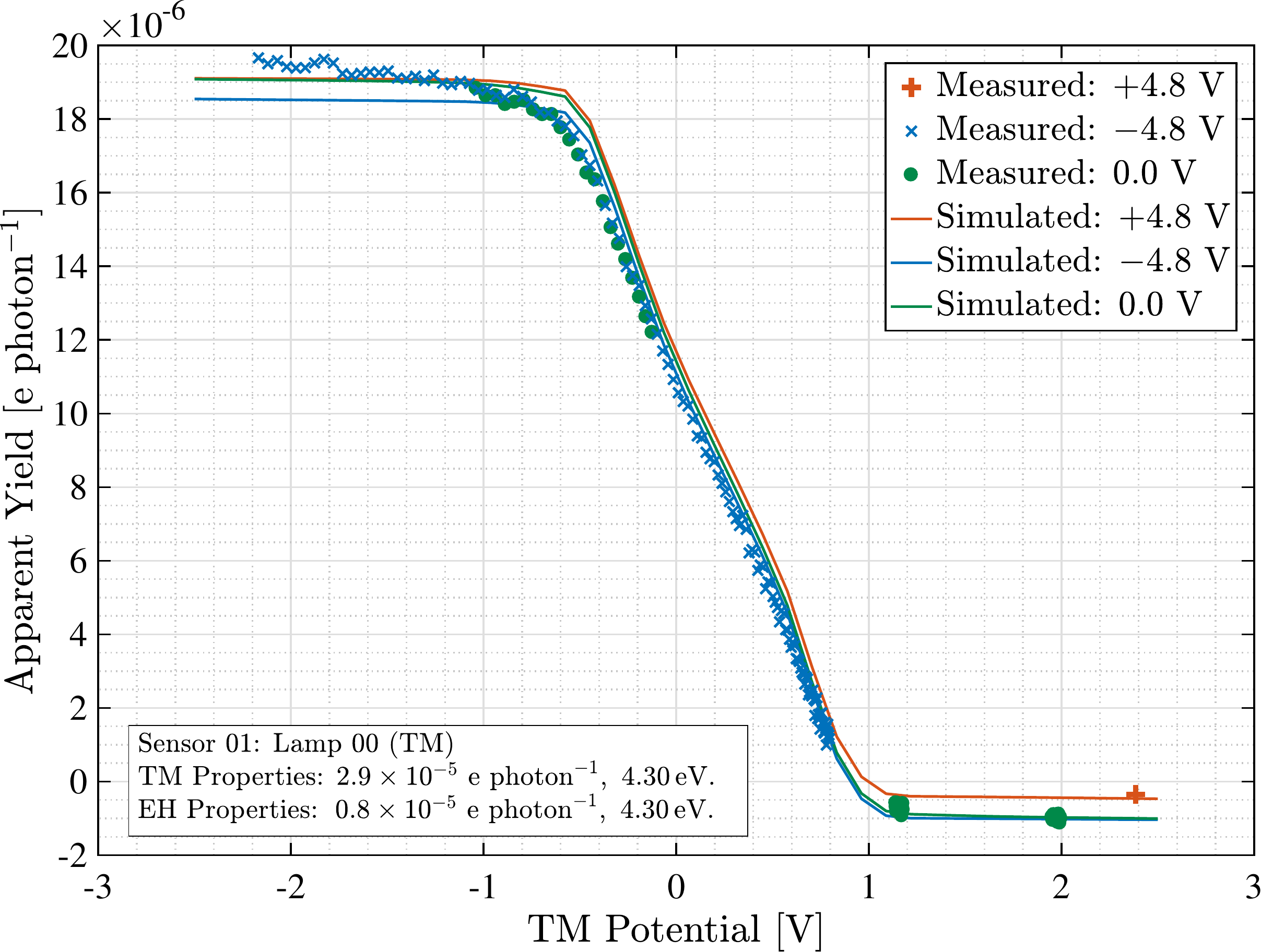}
\includegraphics[width=0.49\textwidth]{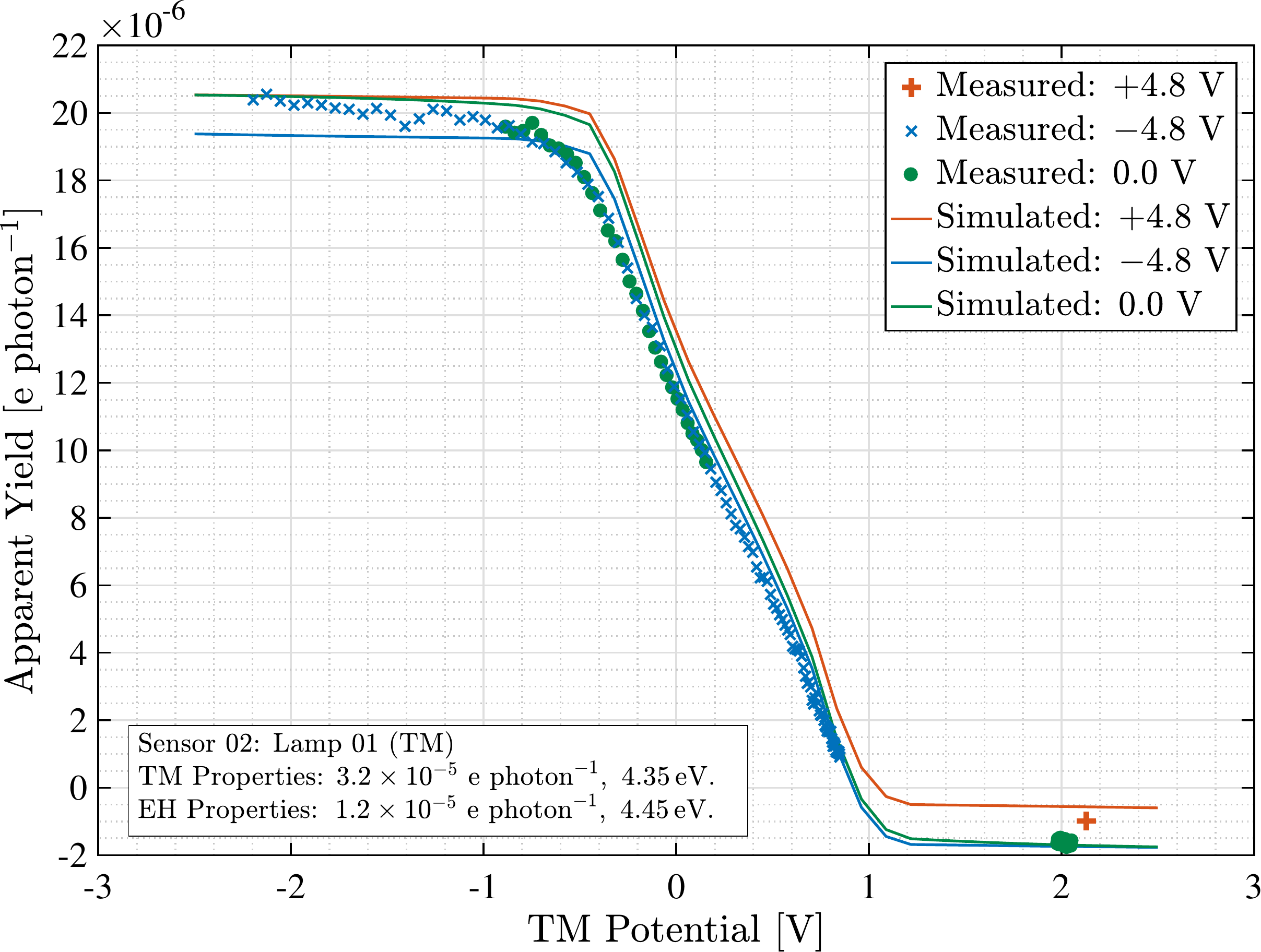}
\includegraphics[width=0.49\textwidth]{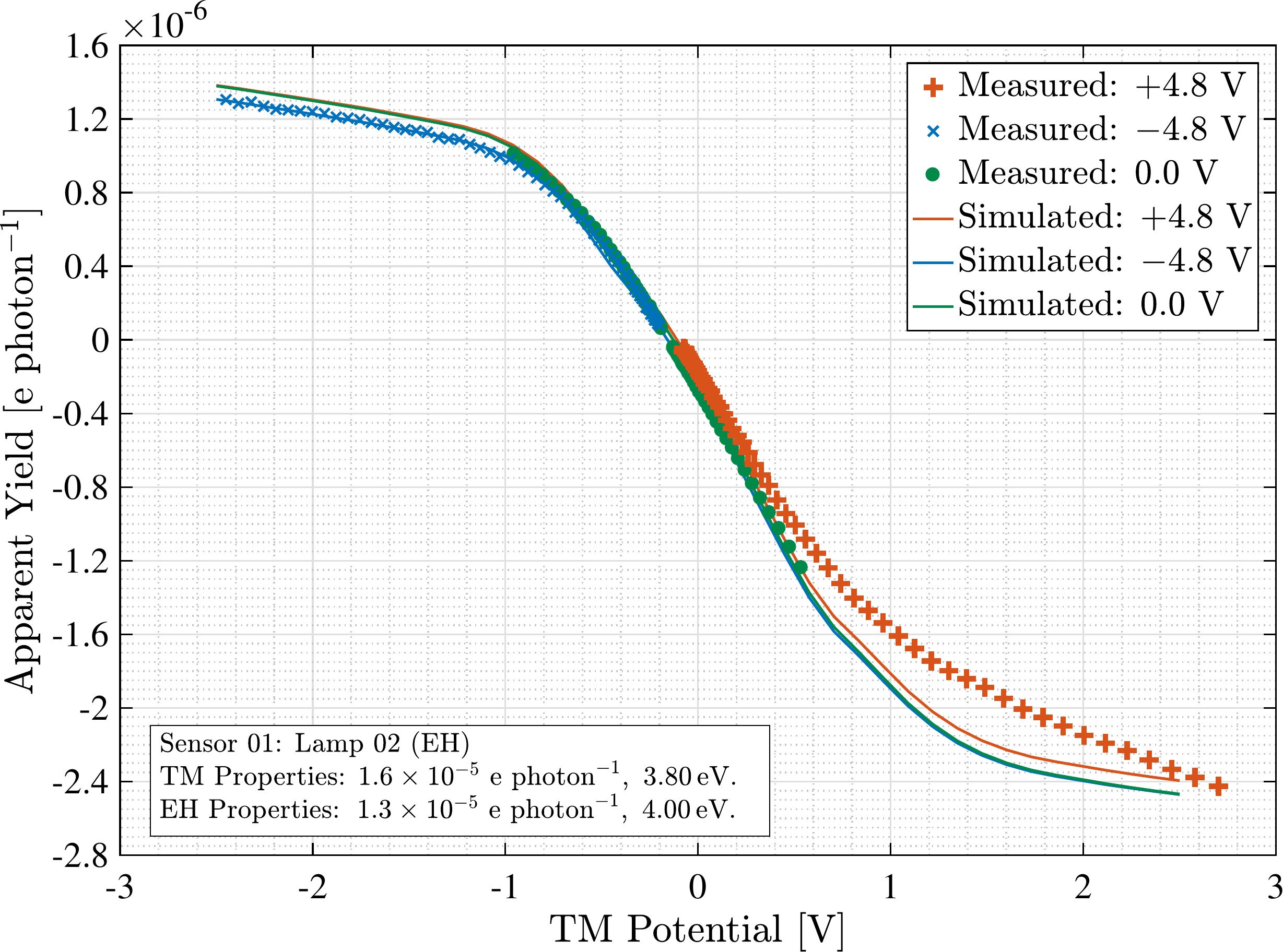}
\includegraphics[width=0.49\textwidth]{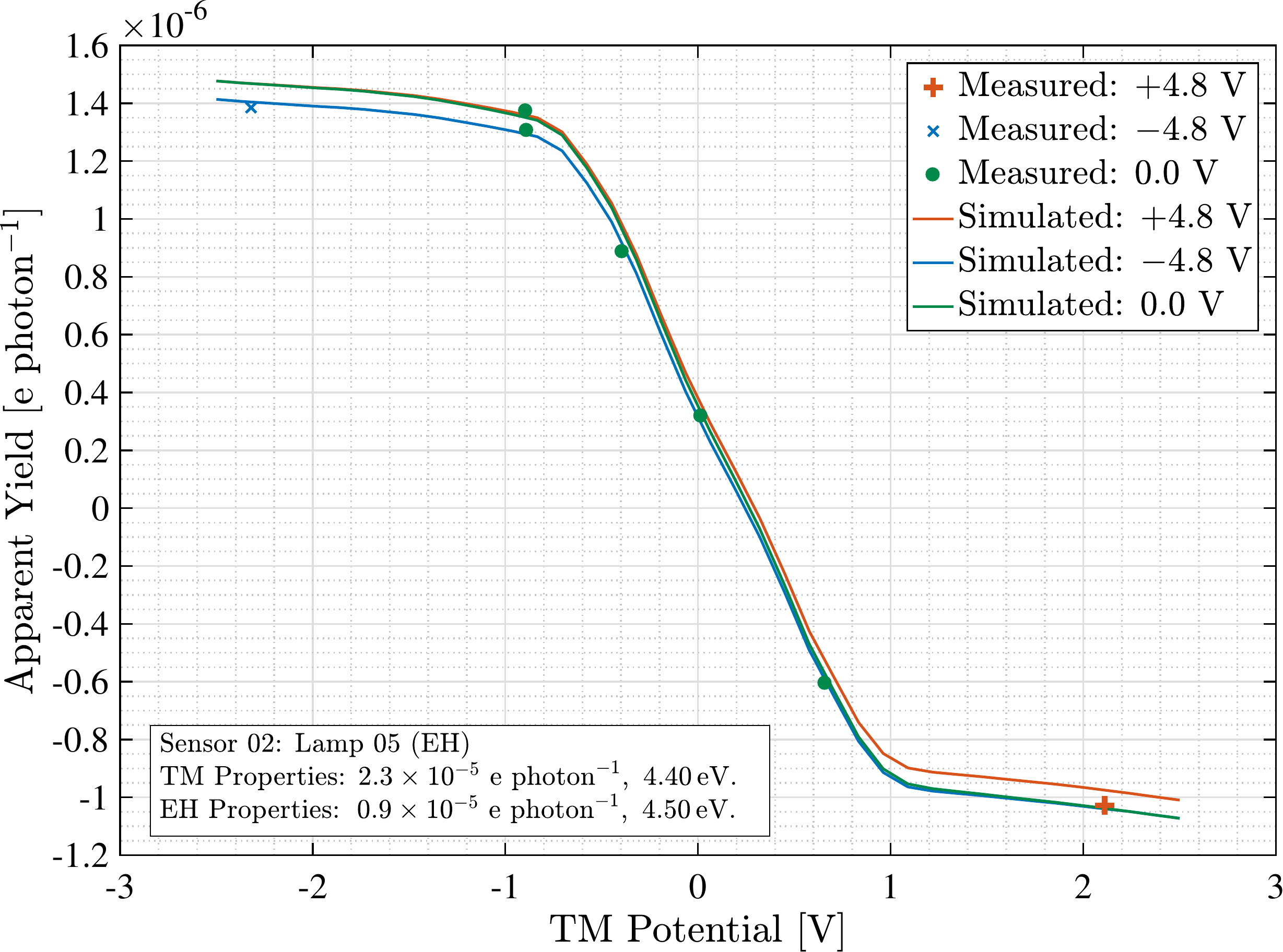}
\includegraphics[width=0.49\textwidth]{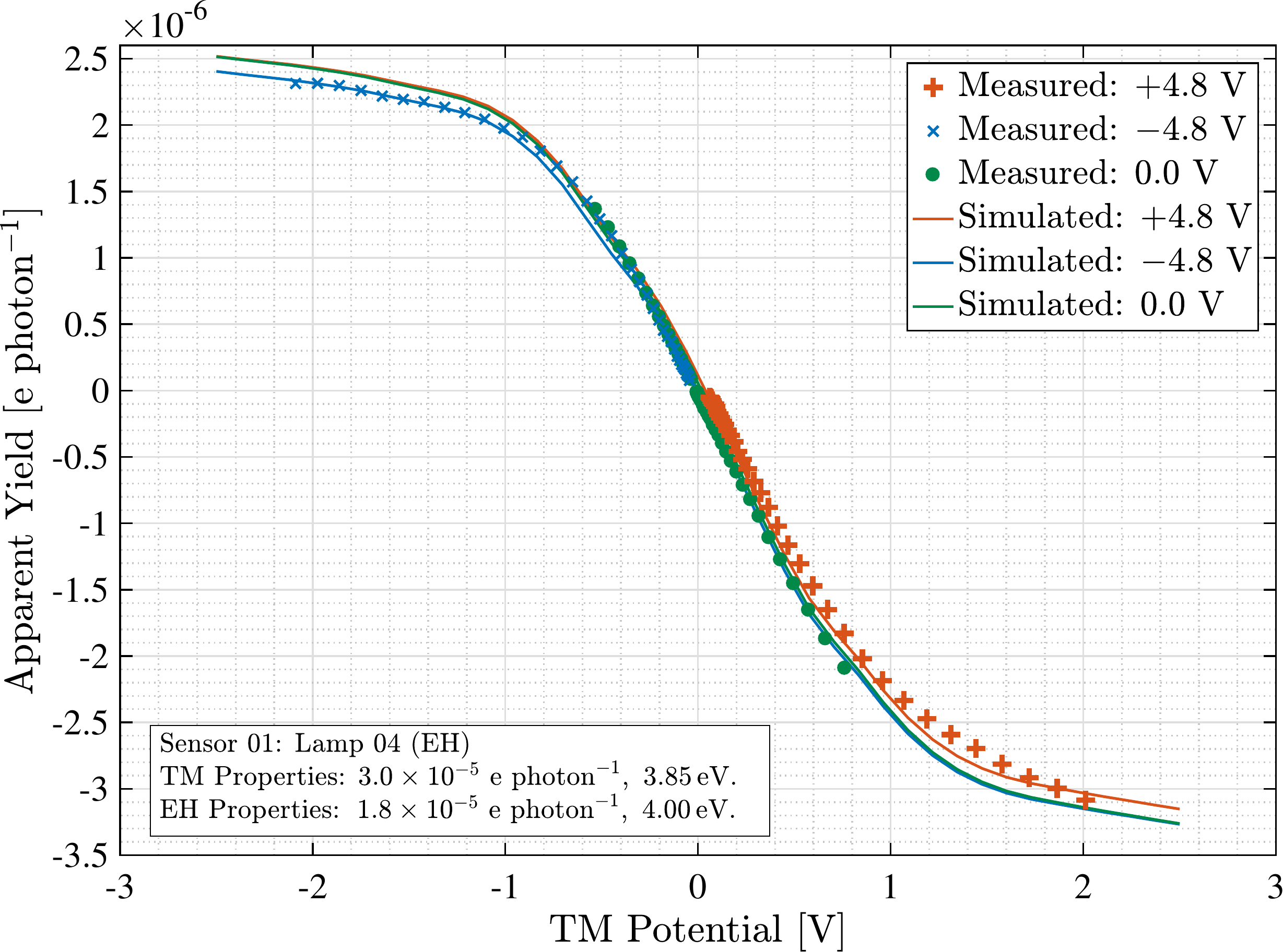}
\includegraphics[width=0.49\textwidth]{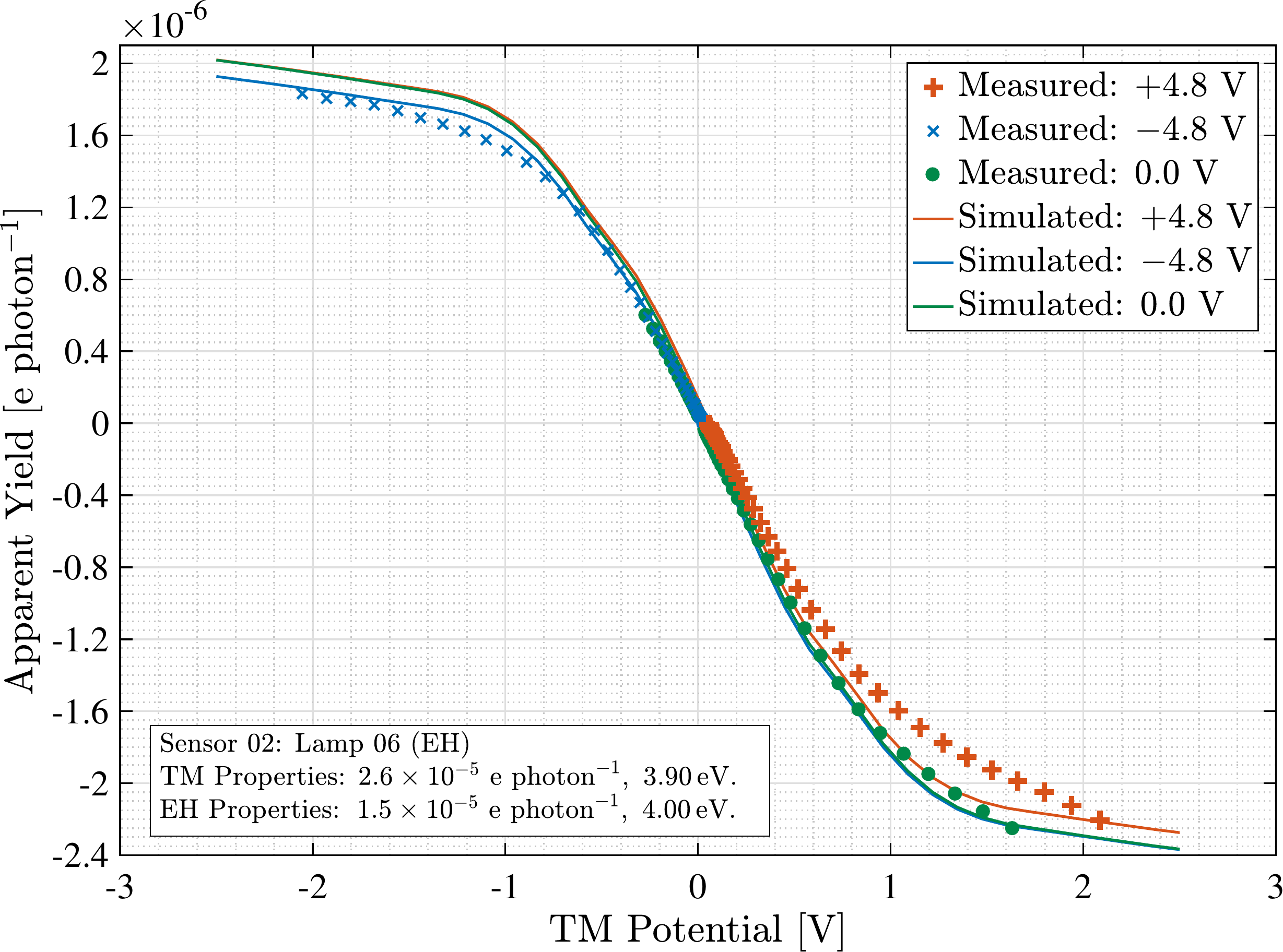}
\caption[Simulated discharge curves.]{\label{fig:simulatedCurves} The six discharge curves measured in-flight, described in Sec. \ref{sec:dischargeCurves}. Those on the left are for sensor 01 while those on the right are for sensor 02. In each case simulated discharge curves have been overlaid. For a single set of photoelectric parameters three discharge curves were simulated for each lamp; one with no DC biases applied, one with $+4.8\,$V applied to all $x$ and $y$ sensing electrodes and one with $-4.8\,$V applied to all $x$ and $y$ sensing electrodes.}
\end{figure*}

To assess the measured discharge curves we first take the output of the ray trace (detailed in Sec. \ref{sec:RayTrace} and tabulated in Appendix \ref{tab:UVabsorption}) as the input to the photoelectron flow model (described in Sec. \ref{sec:PEflow}). The work function and quantum yield parameters of both the test mass and electrode housing surfaces are then varied independently for each curve while checking for consistency with the data. Although the model allows the photoelectric properties to be specified for individual surfaces (each electrode, test mass face, corner dome, etc.) we only considered average properties for either the test mass or housing to avoid arbitrary tuning. Despite this simplification, the model is able to capture many of the features observed in the measured curves with the results shown in Figure \ref{fig:simulatedCurves}.

For each discharge curve there were four free parameters; the average test mass quantum yield, the average housing quantum yield, the average test mass work function and the average housing work function. It was found that the average quantum yields were heavily constrained by the data with the values for the test mass and housing also being independent of each other. Essentially, the quantum yields have to be compatible with the saturation levels at either extreme which typically leads to uncertainties of less than 10\%. Prior measurements on ground had always found the work function of gold (previously exposed to air) to lie between 3.5-4.7\,eV \cite{Hollington2011}, therefore only values in this range were considered. The interplay between the two opposing work functions (that define the photoelectron energy distributions) determine the shape and position of the transition between the two saturation levels and it was found there was some degeneracy in pairs of work functions that gave good results. Nevertheless, our primary concern were the quantum yields and the exact work function values had little influence on these.

The simulated discharge curves generally capture the shape of the measurements very well, particularly at positive yields for lamps aimed at the housing (02, 04, 05, 06). However, the model struggles to accurately reproduce the shifts in the negative yields for lamps 02 and 06 when $+4.8\,$V was applied. One possible explanation is that in these cases the yield from the electrode regions were higher than the average value causing a more prominent shift when the DC bias was applied. Tests have shown that the model is capable of reproducing the observed behaviour by increasing the yield in the electrode regions and proportionally decreasing it elsewhere, thus keeping the same average value. However, there is no way of distinguishing between a single electrode having a much higher yield or several being slightly higher. As such we have stuck to a single average value, though it produces a poorer fit in these regions. It should be emphasised here that the smaller shift observed for lamp 04 and larger shifts for lamps 00 and 01 are well captured by a single average quantum yield. This observation increases confidence that the amount of light predicted to be absorbed in these regions by the ray trace is correct as it needs to be compatible with the average quantum yield. The model also has trouble reproducing the gradually increasing positive saturation levels for the test mass facing lamps, particularly lamp 00. While there is a small gradient in the simulated saturation level it is not as prominent as that observed. Again, tests suggest that by varying the yield in different regions but keeping the average similar the model can reproduce the behaviour but there is no way of distinguishing between several arbitrary ways of tuning.

\begin{figure}[]
\centering
\includegraphics[width=0.5\textwidth]{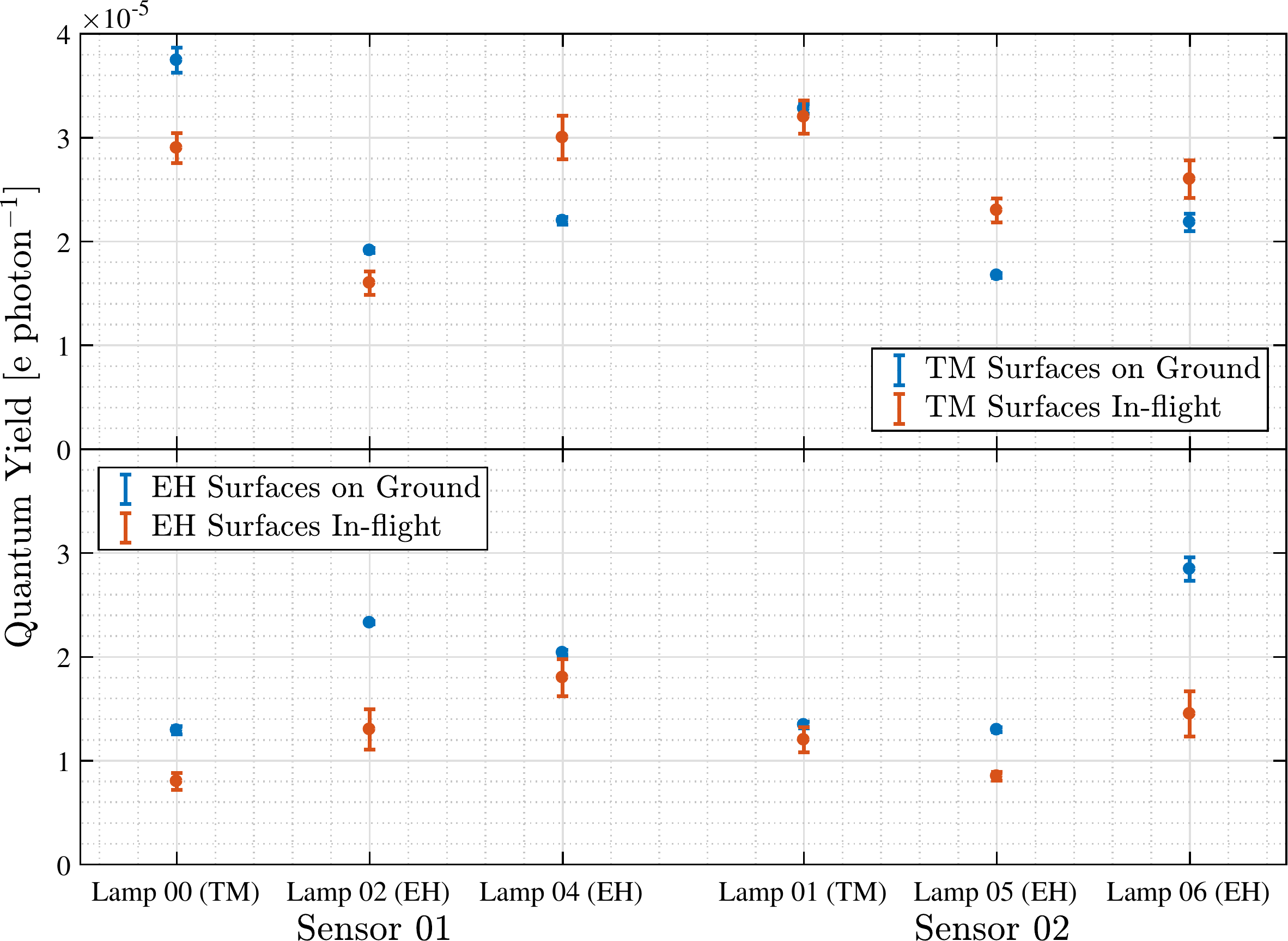}
\caption[Estimated average quantum yields.]{\label{fig:estQYs} The average quantum yields from the test mass surfaces (TM, top panel) and the electrode housing surfaces (EH, bottom panel) for all six lamp illuminations. For each lamp the primary surface under illumination is specified in brackets. Quantum yield estimates are shown based on measurements made after bake-out on ground (Table \ref{tab:flightAYground}) as well as those made in-flight (Figure \ref{fig:simulatedCurves}). Note that there are additional systematic uncertainties at the 10-20\% level on the in-flight estimates due to the UV photodiode calibration.}
\end{figure}

We now have estimates for the average quantum yields of each surface measured in-flight. To allow comparison, the apparent yields measured on ground also need to be converted to average quantum yields. For the ground measurements only the illumination ratios obtained from the ray trace are required as only the discharge curve saturation levels were measured and different bias voltages were not applied to individual electrodes. However, as explained in Sec. \ref{sec:Gmeasure} the measurements made on ground had several subtle differences compared to those made in-flight. As such the ray trace was re-run with appropriate ISUK light distributions (unlike flight, these were identical for the six lamps) and with the central caging plungers engaged. For lamps 00 and 01 which primarily illuminated the test mass, 66.5\% of the total light injected was absorbed by test mass surfaces while 14.8\% was absorbed by electrode housing surfaces. For the four lamps aimed at the housing, 12.9\% of the total light injected was absorbed by test mass surfaces while 18.0\% was absorbed by electrode housing surfaces. As discussed in Sec. \ref{sec:Sim}, the remainder of the light in both cases was absorbed by surfaces unlikely to influence discharging and was therefore ignored. Given the fraction of total light absorbed by each surface type, the apparent yields measured on ground (Table \ref{tab:flightAYground}) can be converted to average quantum yields. A comparison between the average quantum yields of each surface measured on ground and in-flight is shown in Figure \ref{fig:estQYs}.

The estimated quantum yields range from about 1 to $4\times10^{-5}\,\textup{e per absorbed photon}$, consistent with typical historical measurements made on individual gold surfaces during hardware development \cite{Hollington2011}. The estimates suggest that for each lamp any changes in the quantum yield between ground and flight were relatively small, differing by a few tens of percent. For the test mass quantum yields the average change between ground and flight was 22\% while for the electrode housing surfaces it was 31\%. In addition to the random uncertainty on the in-flight quantum yield estimates there are significant systematic uncertainties, approximately at the 10-20\% level, related to the on-ground calibration of the UV power emitted. For a particular lamp this systematic uncertainty in the number of photons reaching the sensor would act to scale the test mass and electrode housing quantum yields in the same way, either increasing or decreasing them together. Looking at Figure \ref{fig:estQYs}, this alone can not explain the changes as several pairs of lamp measurements did not change in the same direction or by the same amount. Another aspect to note is that due to a slightly different illumination and a change in central plunger position, the exact same areas were not illuminated on ground and in-flight. Therefore even if the system had remained completely stable the measured quantum yields would not be expected to be identical as slightly different areas were being probed in each case.

While it was previously shown that the photoelectric properties were stable once in space, they do not appear to have become more uniform as the level of scatter on ground and in-flight is comparable. The average level of the quantum yields for each surface type was also little changed. The average quantum yield (and standard deviation) for the test mass surfaces went from $(2.5\pm0.3)\times10^{-5}\,\textup{e per absorbed photon}$ on ground to $(2.6\pm0.3)\times10^{-5}\,\textup{e per absorbed photon}$ in-flight while the electrode housing surfaces went from $(1.9\pm0.3)\times10^{-5}\,\textup{e per absorbed photon}$ to $(1.2\pm0.2)\times10^{-5}\,\textup{e per absorbed photon}$. The average values also highlight that the quantum yield from the test mass surfaces appeared to be systematically higher than those from the housing, both on ground and in-flight. The possibility that nominally identical gold coated surfaces should differ in this way will be discussed in the context of preparing for the LISA mission.

\section{Toward LISA}

Despite its overall success, measurements made during development, testing and in-flight for the Pathfinder discharge system have identified three key issues that face the LISA system. The first relates to the significant variability in the photoelectric properties of air exposed gold surfaces, when illuminated with 253.7\,nm UV light. Independent of any modelling, the apparent yields measured on ground with the flight sensor had coefficients of variation up to 30\% for groups of measurements with nominally identical illuminations. Modelling the system also allowed the intrinsic quantum yields to be extracted and found that for nominally identical gold surfaces, stored and handled in controlled conditions and undergoing a 24 hour bake-out at $115\,^{\circ}\textup{C}$, the quantum yield varied from 1 to $4\times10^{-5}\,\textup{e per absorbed photon}$. Similar levels of variability have been observed previously and have been interpreted as an extreme sensitivity to the state of the surface \cite{Hechenblaikner2012}. Such variability makes it impossible to accurately predict the final performance of the discharge system and while the ability to apply DC biases makes it very unlikely that fast discharge would be prevented, continuous discharging (relying on bi-polar transfer with no DC bias) could be impractical, as was the case on Pathfinder with three of the four electrode housing lamps. While an improved understanding of how surface contamination affects the quantum yield would certainly be beneficial, in real terms there is limited scope for what could be done differently for LISA. The surfaces will inevitably be exposed to air during integration and strict storage and handling procedures were already in place for Pathfinder. One area of investigation that could prove fruitful is the relationship between bake-out temperature and duration and how they relate to the quantum yield variability.

Alongside the variability, the quantum yield of the test mass surfaces as extracted from our model appear to be systematically higher than that of the housing by around a factor of 2. While an overlooked systematic error in the modelling is always possible it is difficult to envisage where such a large discrepancy could originate. If instead the asymmetry has a physical origin seeking an explanation before moving toward LISA will be important in order to consolidate confidence in the design, production and integration of the sensor. Possible overlooked differences between the test mass and housing gold surfaces include the underlying substrate, the surface roughness and the relative temperatures when cooling after bake-out which could lead to contaminates being redistributed preferentially on one of the surfaces.

The affect of both variability and asymmetry in the quantum yields was exacerbated on Pathfinder by a significant amount of light being lost within recesses when the housing surfaces were illuminated. Combined with contact potentials preventing transport of the low energy photoelectrons, this effectively meant 70\% of the light was wasted when the housing was illuminated. This left the ratio of useful absorbed light between the housing and test mass at just 2.1 for direct housing illumination in-flight. This compared to a ratio of useful absorbed light between the test mass and housing of 4.6 for direct test mass illumination in-flight. Broadly speaking, for test mass illumination the apparent yield from the test mass would still dominate even if the quantum yield was around a factor 4 higher for the housing than for the test mass. For housing illumination, only around a factor 2 difference can be accommodated, which was coincidentally similar to the systematic difference in the observed test mass and electrode housing quantum yields. Though it could of course bring unforeseen risks, there is the possibility that the illumination scheme for LISA could be changed. For example, initial tests suggest that had the lamp 02 ISUK been angled at $8^{\circ}$ rather than $20^{\circ}$, the absorption percentages in-flight would have been 47.7\% for the electrode housing (as opposed to 20.5\% for Pathfinder), 15.2\% for the test mass (9.9\% for Pathfinder), with just 37.0\% wasted, increasing the ratio of useful light to 3.2. However, a detailed investigation would be required to assess the many illumination alternatives and whether such changes were feasible from an engineering perspective. More importantly, there is the possibility that any changes could have unintended consequences elsewhere and there is therefore a strong argument to keep things the same, particularly given the success of the Pathfinder system.

Due to mass, power consumption and lifetime advantages, the LISA CMS will almost certainly be based on UVLEDs rather than mercury lamps \cite{Sun2006, Sun2009, Hollington2015}. While changing the light source will not automatically address the issues associated with variable and asymmetric quantum yields, it could offer ways of mitigating against them. The use of UVLEDs opens up the possibility of driving them in a pulsed mode while synchronising their output to the AC voltages present within the sensor \cite{Sun2006, Ziegler2014, Hollington2017}. Conceptually, the sensor would only be illuminated while the AC voltages were at levels conducive to the desired direction of discharge. Such a scheme should offer an increased dynamic range and robustness against variable and asymmetric quantum yields but would come at the cost of increased complexity and potentially a new set of unknown challenges.

%%%%%%%%%%%%%%%%%%%%%%%%%%%%%%%%%%%%%%%%%%%%%%%%%%

\section{Conclusions}

The Pathfinder CMS operated successfully over the entire mission from March 2016 to July 2017. It had to counter average cosmic ray charging rates from $+21\,\textup{es}^{-1}$ to $+36\,\textup{es}^{-1}$ as well as post-release residual charge that produced test mass potentials as high $-421$\,mV. During the mission over 50 fast discharges, 2 continuous discharging runs and several dedicated discharging tests were performed. The six mercury lamps were turned on a combined total of 418 times without fail and were operated for over 421 hours. The routine fast discharges carried out during station keeping kept the test mass charge within 100\,mV of zero during all primary test mass acceleration runs and continuous discharge tests also demonstrated that the test mass charge could be maintained within 10\,mV of zero for days, likely weeks.

Alongside the production of the physical hardware, a two part software model was developed at Imperial College London to simulate the overall discharging behaviour. By modelling the entire sensor geometry and reflective properties of individual surfaces, the ray trace determined how much UV light was absorbed and where. This information could then be fed into the photoelectron flow model which approximated the system as a collection of parallel plates, greatly simplifying the electric field geometries. Taking into account the many AC and DC voltages within the sensor, as well as the photoelectron energy distributions, the surface quantum yields and work functions were then estimated from the original apparent yield measurements. It was found that this relatively simple model was able to reproduce most of the observed behaviour with shortfalls due to a desire to limit the number of free parameters rather than an intrinsic limitation in the simulation. While a more rigorous treatment of the electric fields would have increased complexity, and therefore runtime of the simulation, it is doubtful a full implementation would have improved results. Instead, it is a lack of knowledge for how the photoelectric properties vary over particular surfaces that ultimately limits the simulation. There is no practical way of directly measuring the photoelectric properties of individual flight surfaces \textit{in situ} and due to their propensity to vary upon re-exposure to air, measurements would be of little use prior to integration. However, the simulation could be improved with direct roughness and reflectivity measurements, particularly for the iridium and gold-platinum surfaces.

The ability to measure each lamp's apparent yield on ground with the flight sensors proved very useful and increased confidence that the discharge system would work successfully in space. There are several ways in which similar measurements for LISA could be improved. Firstly, the optical chains (including any attenuators) need to be identical to those used in-flight as they affect the distribution of UV light emitted from the ISUKs and therefore the area illuminated within the sensor. Secondly, a lamp's entire discharge curve should be measured rather than just the saturation levels, ideally with AC voltages applied to electrodes that are representative of those experienced in-flight. Less extrapolation would then be needed to predict the final in-flight performance and also allow a better test of simulation. Finally, and if time permitted, different DC biases could be applied to individual electrodes to assess how different regions contribute to the total apparent yield, either through absorbing a lot of light or having a high quantum yield. Not only would this provide a more stringent test of simulated predictions but also allow fast discharge voltage schemes to be adjusted to give optimal performance in-flight.

Measurements made in-flight showed behaviour for each lamp that was qualitatively similar to that observed on ground and further analysis using the full simulation showed reasonable quantitative agreement. Despite the estimated internal sensor pressure being around a millibar for over 15 months prior to the launch, the test mass quantum yields changed by an average of only 22\% while for the electrode housing surfaces it was 31\%. A separate analysis using measurements made during routine fast discharges showed that the apparent yields for lamps 02 and 06 were stable within the measurement uncertainty 140 days after the system was vented to space and consistent with another measurement made during commissioning just 25 days after venting. The measured apparent yields were stable with time under vacuum, UV irradiation from discharging, cosmic ray irradiation and a temperature reduction of about $11\,^{\circ}\textup{C}$. This analysis also demonstrated that the optical chain transmission between each lamp and the sensors did not degrade over the course of the mission.

Moving toward development of the LISA discharge system the issues of sub-optimal illumination of the housing, variability in the quantum yield of air exposed gold surfaces and the possible asymmetry in test mass and housing yields will need to be addressed. While all the lamps were capable of fast discharge these issues conspired to make all but one of the four housing illuminations less than ideal for automated continuous discharging as they lacked a significant negative apparent yield at a test mass potential of 0\,V. Given the lessons learned from the Pathfinder CMS there are essentially two options for the LISA system if continuous discharge capability is deemed desirable. The first would be to understand and ultimately reduce the quantum yield variability and asymmetry allowing for a simple system similar to Pathfinder, though likely replacing the mercury lamps with UVLEDs. Alternatively, it could be accepted that these factors cannot be well controlled in a practical way and instead design a more complicated modulated CMS that can mitigate against their affect. Given the early stage of LISA development, it is likely that both strategies will continue to be explored.

%%%%%%%%%%%%%%%%%%%%%%%%%%%%%%%%%%%%%%%%%%%%%%%%%%

\section*{Acknowledgements}

This work has been made possible by the LISA Pathfinder mission, which is part of the space-science program of the European Space Agency. The French contribution has been supported by CNES (Accord Specific de Projet No. CNES 1316634/CNRS 103747), the CNRS, the Observatoire de Paris and the University Paris-Diderot. E.~Plagnol and H.~Inchausp\'e would also like to acknowledge the financial support of the UnivEarthS Labex program at Sorbonne Paris Cit (Grants No. ANR-10-LABX-0023 and No. ANR-11-IDEX-0005-02). The Albert-Einstein-Institut acknowledges the support of the German Space Agency, DLR. The work is supported by the Federal Ministry for Economic Affairs and Energy based on a resolution of the German Bundestag (Grants No. FKZ 50OQ0501 and No. FKZ 50OQ1601). The Italian contribution has been supported by Agenzia Spaziale Italiana and Instituto Nazionale di Fisica Nucleare. The Spanish contribution has been supported by Contracts No. AYA2010-15709 (MICINN), No. ESP2013-47637-P, and No. ESP2015-67234-P (MINECO). M.~Nofrarias acknowledges support from Fundacion General CSIC Programa ComFuturo). F.~Rivas acknowledges support from a Formacin de Personal Investigador (MINECO) contract. The Swiss contribution acknowledges the support of the Swiss Space Office (SSO) via the PRODEX Programme of ESA. L.~Ferraioli acknowledges the support of the Swiss National Science Foundation. The UK groups wish to acknowledge support from the United Kingdom Space Agency (UKSA), the University of Glasgow, the University of Birmingham, Imperial College London, and the Scottish Universities Physics Alliance (SUPA). N.~Korsakova would like to acknowledge the support of the Newton International Fellowship from the Royal Society. J.\,I.~Thorpe and J.~Slutsky acknowledge the support of the U.S. National Aeronautics and Space Administration (NASA).

%%%%%%%%%%%%%%%%%%%%%%%%%%%%%%%%%%%%%%%%%%%%%%%%%%

%\section*{References}

\bibliographystyle{unsrt}
\bibliography{bibliography}

\begin{thebibliography}{10}

\bibitem{Amaro2017}
P.~Amaro-Seoane et~al.
\newblock {Laser Interferometer Space Antenna}.
\newblock {\em arXiv preprint arXiv:1702.00786v3}, 2017.

\bibitem{Armano2016}
M.~Armano et~al.
\newblock {Sub-Femto-$g$ Free Fall for Space-Based Gravitational Wave
  Observatories: LISA Pathfinder Results}.
\newblock {\em Phys. Rev. Lett.}, 116:231101, Jun 2016.

\bibitem{Armano2018a}
M.~Armano et~al.
\newblock {Beyond the Required LISA Free-Fall Performance: New LISA Pathfinder
  Results Down to $20\text{ }\text{ }\ensuremath{\mu}\mathrm{Hz}$}.
\newblock {\em Phys. Rev. Lett.}, 120:061101, Feb 2018.

\bibitem{Shaul2005}
D.~N.~A. Shaul, H.~M. Ara{\'u}jo, G.~K. Rochester, T.~J. Sumner, and P.~J.
  Wass.
\newblock {Evaluation of Disturbances Due to Test Mass Charging for LISA}.
\newblock {\em Classical and Quantum Gravity}, 22(10):S297, 2005.

\bibitem{Armano2017}
M.~Armano et~al.
\newblock {Charge-Induced Force Noise on Free-Falling Test Masses: Results from
  LISA Pathfinder}.
\newblock {\em Phys. Rev. Lett.}, 118:171101, Apr 2017.

\bibitem{Araujo2005}
H.~M. Ara{\'u}jo, P.~Wass, D.~Shaul, G.~Rochester, and T.~J. Sumner.
\newblock {Detailed Calculation of Test-Mass Charging in the LISA Mission}.
\newblock {\em Astroparticle Physics}, 22(5-6):451--469, 2005.

\bibitem{Wass2005}
P.~J. Wass, H.~M. Ara{\'u}jo, D.~N.~A. Shaul, and T.~J. Sumner.
\newblock {Test-Mass Charging Simulations for the LISA Pathfinder Mission}.
\newblock {\em Classical and Quantum Gravity}, 22(10):S311, 2005.

\bibitem{Jafry1997}
Y.~Jafry and T.~J. Sumner.
\newblock {Electrostatic Charging of the LISA Proof Masses}.
\newblock {\em Classical and Quantum Gravity}, 14(6):1567, 1997.

\bibitem{Armano2018b}
M.~Armano et~al.
\newblock {Measuring the Galactic Cosmic Ray Flux with the LISA Pathfinder
  Radiation Monitor}.
\newblock {\em Astroparticle Physics}, 98:28--37, Mar 2018.

\bibitem{Armano2018c}
M.~Armano et~al.
\newblock {Characteristics and Energy Dependence of Recurrent Galactic
  Cosmic-Ray Flux Depressions and of a Forbush Decrease with LISA Pathfinder}.
\newblock {\em The Astrophysical Journal}, 854(2):113, 2018.

\bibitem{Reigber2002}
C.~Reigber, H.~Luhr, and Schwintzer P.
\newblock {CHAMP Mission Status}.
\newblock {\em Advances in Space Research}, 30(2):129--134, 2002.

\bibitem{Tapley2004}
B.~D. Tapley, S.~Bettadpur, M.~Watkins, and C.~Reigber.
\newblock {The Gravity Recovery and Climate Experiment: Mission Overview and
  Early Results}.
\newblock {\em Geophysical Research Letters}, 31(9), 2004.

\bibitem{Rummel2011}
R.~Rummel, W.~Yi, and C.~Stummer.
\newblock {GOCE Gravitational Gradiometry}.
\newblock {\em Journal of Geodesy}, 85(11):777--790, 2011.

\bibitem{Touboul2001}
P.~Touboul and M.~Rodrigues.
\newblock {The MICROSCOPE Space Mission}.
\newblock {\em Classical and Quantum Gravity}, 18(13):2487, 2001.

\bibitem{Buchman1995}
Saps Buchman, Theodore Quinn, G.~M. Keiser, Dale Gill, and T.~J. Sumner.
\newblock {Charge Measurement and Control for the Gravity Probe B Gyroscopes}.
\newblock {\em Review of Scientific Instruments}, 66(1):120--129, 1995.

\bibitem{Weber2003}
William~J. Weber, Daniele Bortoluzzi, Antonella Cavalleri, Ludovico Carbone,
  Mauro Da~Lio, Rita Dolesi, Giorgio Fontana, C.~D. Hoyle, Mauro Hueller, and
  Stefano Vitale.
\newblock {Position Sensors for Flight Testing of LISA Drag-Free Control}.
\newblock In {\em Astronomical Telescopes and Instrumentation}, pages 31--42.
  International Society for Optics and Photonics, 2003.

\bibitem{Bortoluzzi2009}
D.~Bortoluzzi, L.~Baglivo, M.~Benedetti, F.~Biral, P.~Bosetti, A.~Cavalleri,
  M.~Da Lio, M.~De Cecco, R.~Dolesi, M.~Lapolla, W.~Weber, and S.~Vitale.
\newblock {LISA Pathfinder Test Mass Injection in Geodesic Motion: Status of
  the On-ground Testing}.
\newblock {\em Classical and Quantum Gravity}, 26(9):094011, 2009.

\bibitem{Audley2011}
H.~Audley et~al.
\newblock {The LISA Pathfinder Interferometry-Hardware and System Testing}.
\newblock {\em Classical and Quantum Gravity}, 28(9):094003, 2011.

\bibitem{Antonucci2012}
F.~Antonucci, A.~Cavalleri, R.~Dolesi, M.~Hueller, D.~Nicolodi, H.~B. Tu,
  S.~Vitale, and W.~J. Weber.
\newblock {Interaction between Stray Electrostatic Fields and a Charged
  Free-Falling Test Mass}.
\newblock {\em Phys. Rev. Lett.}, 108:181101, Apr 2012.

\bibitem{Huber1966}
E.~E. Huber~Jr.
\newblock {The Effect of Mercury Contamination on the Work Function of Gold}.
\newblock {\em Applied Physics Letters}, 8(7):169--171, 1966.

\bibitem{Saville1995}
G.~F. Saville, P.~M. Platzman, G.~Brandes, R.~Ruel, and R.~L. Willett.
\newblock {Feasibility Study of Photocathode Electron Projection Lithography}.
\newblock {\em Journal of Vacuum Science and Technology}, 13(6):2184--2188,
  1995.

\bibitem{Hechenblaikner2012}
Gerald Hechenblaikner, Tobias Ziegler, Indro Biswas, Christoph Seibel, Mathias
  Schulze, Nico Brandt, Achim Schöll, Patrick Bergner, and Friedrich~T.
  Reinert.
\newblock {Energy Distribution and Quantum Yield for Photoemission from
  Air-Contaminated Gold Surfaces Under Ultraviolet Illumination Close to the
  Threshold}.
\newblock {\em Journal of Applied Physics}, 111(12):124914, 2012.

\bibitem{Schulte2009}
M.~O. Schulte, D.~N.~A. Shaul, D.~Hollington, S.~Waschke, T.~J. Sumner, P.~J.
  Wass, L.~Pasquali, and S.~Nannarone.
\newblock {Inertial Sensor Surface Properties for LISA Pathfinder and their
  Effect on Test Mass Discharging}.
\newblock {\em Classical and Quantum Gravity}, 26(9):094008, 2009.

\bibitem{Haynes2010}
W.~M. Haynes and D.~R. Lide.
\newblock {\em {CRC Handbook of Chemistry and Physics}}.
\newblock Taylor \& Francis Group, 2010.

\bibitem{Hollington2011}
D.~Hollington.
\newblock {\em {The Charge Management System for LISA and LISA Pathfinder}}.
\newblock PhD thesis, Imperial College London, 2011.

\bibitem{Cavalleri2009}
A.~Cavalleri, G.~Ciani, R.~Dolesi, M.~Hueller, D.~Nicolodi, D.~Tombolato, P.~J.
  Wass, W.~J. Weber, S.~Vitale, and L.~Carbone.
\newblock Direct force measurements for testing the lisa pathfinder
  gravitational reference sensor.
\newblock {\em Classical and Quantum Gravity}, 26(9):094012, 2009.

\bibitem{Antonucci2010}
F.~Antonucci, A.~Cavalleri, R.~Dolesi, A.~Hueller, S.~Perreca, S.~Vitale, and
  W.~J. Weber.
\newblock {Test-Mass Discharging Test Campaign on TM3 with the Four-Test-Mass
  Torsion Pendulum}.
\newblock {S2-UTN-TN-3080 Issue 1.0}, {University of Trento}, 2010.

\bibitem{Wass2006}
P.~J. Wass, L.~Carbone, A.~Cavalleri, G.~Ciani, R.~Dolesi, M.~Hueller,
  G.~Rochester, M.~Schulte, T.~Sumner, D.~Tombolato, C.~Trenkel, S.~Vitale, and
  W.~Weber.
\newblock Testing of the uv discharge system for lisa pathfinder.
\newblock {\em AIP Conference Proceedings}, 873(1):220--224, 2006.

\bibitem{Tombolato2008}
D.~Tombolato.
\newblock {\em {A Laboratory Study of Force Disturbances for the LISA Free Fall
  Demonstration Mission}}.
\newblock PhD thesis, Universit{\`a} di Trento, 2008.

\bibitem{Ziegler2014}
Tobias Ziegler, Patrick Bergner, Gerald Hechenblaikner, Nico Brandt, and Walter
  Fichter.
\newblock {Modeling and Performance of Contact-Free Discharge Systems for Space
  Inertial Sensors}.
\newblock {\em IEEE Transactions on Aerospace and Electronic Systems},
  50(2):1493--1510, 2014.

\bibitem{Agostinelli2003}
S.~Agostinelli et~al.
\newblock {Geant4 — A Simulation Toolkit}.
\newblock {\em Nuclear Instruments and Methods in Physics Research Section A:
  Accelerators, Spectrometers, Detectors and Associated Equipment},
  506(3):250--303, 2003.

\bibitem{Allison2006}
J.~Allison et~al.
\newblock {Geant4 Developments and Applications}.
\newblock {\em IEEE Transactions on Nuclear Science}, 53(1):270--278, Feb 2006.

\bibitem{Allison2016}
J.~Allison et~al.
\newblock {Recent Developments in Geant4}.
\newblock {\em Nuclear Instruments and Methods in Physics Research Section A:
  Accelerators, Spectrometers, Detectors and Associated Equipment},
  835:186--225, 2016.

\bibitem{Johnson1972}
P.~B. Johnson and R.~W. Christy.
\newblock {Optical Constants of the Noble Metals}.
\newblock {\em Phys. Rev. B}, 6(12):4370--4379, Dec 1972.

\bibitem{Palik1998}
E.~D. Palik and G.~Ghosh.
\newblock {\em {Handbook of Optical Constants of Solids}}.
\newblock Academic Press, 1998.

\bibitem{Torrance1967}
K.~E. Torrance and E.~M. Sparrow.
\newblock {Theory for Off-Specular Reflection From Roughened Surfaces}.
\newblock {\em J. Opt. Soc. Am.}, 57(9):1105--1114, Sep 1967.

\bibitem{Weber2007}
W.~J. Weber, L.~Carbone, A.~Cavalleri, R.~Dolesi, C.~D. Hoyle, M.~Hueller, and
  S.~Vitale.
\newblock Possibilities for measurement and compensation of stray dc electric
  fields acting on drag-free test masses.
\newblock {\em Advances in Space Research}, 39(2):213 -- 218, 2007.

\bibitem{Speake1996}
C.~C. Speake.
\newblock {Forces and Force Gradients Due to Patch Fields and Contact-Potential
  Differences}.
\newblock {\em Classical and Quantum Gravity}, 13(11A):A291, 1996.

\bibitem{Sun2006}
K-S. Sun, S.~Buchman, B.~Allard, S.~Williams, and R.~L. Byer.
\newblock {LED Deep UV Source for Charge Management of Gravitational Reference
  Sensors}.
\newblock {\em Classical and Quantum Gravity}, 23(8):S141--S150, 2006.

\bibitem{Sun2009}
K-S. Sun, N.~Leindecker, S.~Higuchi, J.~Goebel, S.~Buchman, and R.~L. Byer.
\newblock {UV LED Operation Lifetime and Radiation Hardness Qualification for
  Space Flights}.
\newblock {\em Journal of Physics: Conference Series}, 154(1):012028, 2009.

\bibitem{Hollington2015}
D.~Hollington, J.~T. Baird, T.~J. Sumner, and P.~J. Wass.
\newblock {Characterising and Testing Deep UV LEDs for Use in Space
  Applications}.
\newblock {\em Classical and Quantum Gravity}, 32(23):235020, 2015.

\bibitem{Hollington2017}
D.~Hollington, J.~T. Baird, T.~J. Sumner, and P.~J. Wass.
\newblock {Lifetime Testing UV LEDs for Use in the LISA Charge Management
  System}.
\newblock {\em Classical and Quantum Gravity}, 34(20):205009, 2017.

\end{thebibliography}

%%%%%%%%%%%%%%%%%%%%%%%%%%%%%%%%%%%%%%%%%%%%%%%%%%

\begin{appendices}

% Reset the figure/table numbering.
\setcounter{figure}{0} \renewcommand{\thefigure}{\Alph{figure}}
\setcounter{table}{0} \renewcommand{\thetable}{\Alph{table}}

\onecolumngrid
\newpage
\section{}

\begin{table}[h!]
\begin{center}
\begin{tabular}{  r | c  c | c  c | c  c | c  c | c  c | c  c | }
		     & \multicolumn{6}{c|}{Sensor 01} & \multicolumn{6}{c|}{Sensor 02}  \\                 
		     & \multicolumn{2}{c|}{Lamp 00 (TM)} & \multicolumn{2}{c|}{Lamp 02 (EH)} & \multicolumn{2}{c|}{Lamp 04 (EH)} & \multicolumn{2}{c|}{Lamp 01 (TM)} & \multicolumn{2}{c|}{Lamp 05 (EH)} & \multicolumn{2}{c|}{Lamp 06 (EH)}  \\
                          & EH             & TM             & EH             & TM            & EH             & TM            & EH             & TM             & EH             & TM            & EH             & TM             \\
\hline
-X Housing                & 0.17           & 0.01           & 1.27  & 0.24 & 0.83           & 0.05          & 2.30  & 12.09 & 0.30           & 0.02          & 1.21  & 0.22  \\
-X Sensing Electrode 01     & 0.00           & 0.00           & 0.21           & 0.09          & 0.00           & 0.00          & 0.14           & 0.13           & 0.00           & 0.00          & 0.18           & 0.09           \\
-X Sensing Electrode 02     & 0.00           & 0.00           & 0.02           & 0.02          & 0.00           & 0.00          & 1.72  & 1.17  & 0.00           & 0.00          & 0.02           & 0.02           \\
+X Housing                & 2.85  & 7.44  & 0.88           & 0.06          & 1.26  & 0.24 & 0.52           & 0.03           & 1.15  & 0.21 & 0.59           & 0.04           \\
+X Sensing Electrode 01     & 0.78  & 0.94  & 0.00           & 0.00          & 0.02           & 0.02          & 0.00           & 0.00           & 0.02           & 0.02          & 0.00           & 0.00           \\
+X Sensing Electrode 02     & 0.07           & 0.07           & 0.00           & 0.00          & 0.20           & 0.09          & 0.00           & 0.00           & 0.18           & 0.09          & 0.00           & 0.00           \\
-Y Housing                & 2.48  & 7.33  & 1.20  & 0.19          & 0.78           & 0.02          & 0.58           & 0.03           & 0.28           & 0.01          & 1.12  & 0.19  \\
-Y Sensing Electrode 01     & 0.66  & 0.63  & 0.16           & 0.07          & 0.00           & 0.00          & 0.00           & 0.00           & 0.00           & 0.00          & 0.13           & 0.06           \\
-Y Sensing Electrode 02     & 0.02           & 0.01           & 0.01           & 0.01          & 0.00           & 0.00          & 0.00           & 0.00           & 0.00           & 0.00          & 0.01           & 0.01           \\
-Y Injection Electrode 00 & 0.09           & 0.07           & 0.05           & 0.04          & 0.00           & 0.00          & 0.00           & 0.00           & 0.00           & 0.00          & 0.05           & 0.04           \\
+Y Housing                & 0.21           & 0.01           & 0.83           & 0.02          & 1.18  & 0.19          & 1.78  & 12.01 & 1.06  & 0.19 & 0.50           & 0.02           \\
+Y Sensing Electrode 01     & 0.00           & 0.00           & 0.00           & 0.00          & 0.15           & 0.06          & 1.86  & 0.99  & 0.11           & 0.05          & 0.00           & 0.00           \\
+Y Sensing Electrode 02     & 0.00           & 0.00           & 0.00           & 0.00          & 0.01           & 0.01          & 0.01           & 0.00           & 0.01           & 0.01          & 0.00           & 0.00           \\
+Y Injection Electrode 00 & 0.00           & 0.00           & 0.00           & 0.00          & 0.05           & 0.04          & 0.14           & 0.12           & 0.06           & 0.04          & 0.00           & 0.00           \\
-Z Housing                & 4.79  & 1.04  & 8.14  & 3.17 & 8.07  & 3.10 & 5.10  & 2.12  & 7.24  & 2.85 & 7.79  & 3.03  \\
-Z Sensing Electrode 01     & 0.01           & 0.00           & 5.19  & 0.91 & 0.09           & 0.13          & 0.43           & 0.97           & 0.03           & 0.05          & 4.24  & 0.94  \\
-Z Sensing Electrode 02     & 0.44           & 0.39           & 0.09           & 0.14          & 4.82  & 0.92 & 0.06           & 0.01           & 2.21  & 1.09 & 0.06           & 0.11           \\
-Z Injection Electrode 01 & 0.25           & 0.04           & 0.85           & 0.77 & 1.36  & 0.23 & 0.33           & 0.09           & 0.47           & 0.06          & 0.56           & 0.60  \\
-Z Injection Electrode 02 & 0.10           & 0.03           & 1.49  & 0.26 & 0.81           & 0.72 & 0.48           & 0.05           & 0.54  & 0.31 & 1.07  & 0.15           \\
+Z Housing                & 0.48           & 0.01           & 0.11           & 0.01          & 0.12           & 0.01          & 0.18           & 0.01           & 0.21           & 0.01          & 0.14           & 0.01           \\
+Z Sensing Electrode 01     & 0.00           & 0.00           & 0.00           & 0.00          & 0.00           & 0.00          & 0.00           & 0.00           & 0.00           & 0.00          & 0.00           & 0.00           \\
+Z Sensing Electrode 02     & 0.00           & 0.00           & 0.00           & 0.00          & 0.00           & 0.00          & 0.00           & 0.00           & 0.00           & 0.00          & 0.00           & 0.00           \\
+Z Injection Electrode 01 & 0.00           & 0.00           & 0.00           & 0.00          & 0.00           & 0.00          & 0.00           & 0.00           & 0.00           & 0.00          & 0.00           & 0.00           \\
+Z Injection Electrode 02 & 0.00           & 0.00           & 0.00           & 0.00          & 0.00           & 0.00          & 0.00           & 0.00           & 0.00           & 0.00          & 0.00           & 0.00           \\
\hline
\hline
All Above Surfaces   & 13.41 & 18.03          & 20.50 & 5.99          & 19.73 & 5.83          & 15.63 & 29.85          & 13.86 & 5.02          & 17.69 & 5.51           \\
Test Mass Corner Domes    & -              & 42.58 & -              & 0.77          & -              & 0.74          & -              & 23.17 & -              & 0.69          & -              & 0.69           \\
Test Mass Corner Regions    & -              & 5.40           & -              & 0.47          & -              & 0.42          & -              & 11.40          & -              & 0.22          & -              & 0.34           \\
Test Mass Central Recess    & -              & 0.06           & -              & 2.62          & -              & 2.37          & -              & 0.08           & -              & 0.91          & -              & 2.06           \\
Central Plungers (Wasted)   & 0.01           & -              & 0.11           & -             & 0.11           & -             & 0.01           & -              & 0.17           & -             & 0.13           & -              \\
Caging Fingers (Wasted)       & 5.64           & -              & 11.80          & -             & 11.61          & -             & 2.97           & -              & 11.08          & -             & 11.26          & -              \\
Housing Recesses (Wasted)       & 14.88          & -              & 57.75 & -             & 59.19 & -             & 16.89          & -              & 68.06 & -             & 62.33 & -              \\
\hline
\hline
Total Useful Light                 & 13.41          & 66.07 & 20.50          & 9.85 & 19.73          & 9.37 & 15.63          & 64.50 & 13.86          & 6.84 & 17.69          & 8.60  \\
\hline
\end{tabular}
\caption[Ray trace results for flight geometry.]{\label{tab:UVabsorption} Percentage of total UV light injected that is absorbed by each region, belonging to either the electrode housing (EH) or test mass (TM). X, Y, Z and +/- denote the different faces of the housing and test mass.}
\end{center}
\end{table}

\end{appendices}

\end{document}